\documentclass[twocolumn]{aastex61}
\usepackage{mathrsfs}

\shortauthors{Pinilla et al.}

\begin{document}

\title{A Multi-Wavelength Analysis of Dust and Gas in the SR 24S Transition Disk}

\correspondingauthor{Paola~Pinilla, Hubble fellow}
\affiliation{Department of Astronomy/Steward Observatory, The University of Arizona, 933 North Cherry Avenue, Tucson, AZ 85721, USA}
\email{pinilla@email.arizona.edu}

\author{P.~Pinilla}
\affiliation{Department of Astronomy/Steward Observatory, The University of Arizona, 933 North Cherry Avenue, Tucson, AZ 85721, USA}

\author{L.~M.~P\'erez}
\affiliation{Max-Planck-Institut f\"{u}r Radioastronomie, Auf dem H\"{u}gel 69, 53121 Bonn, Germany}

\author{S.~Andrews}
\affiliation{Harvard-Smithsonian Center for Astrophysics, 60 Garden Street, Cambridge, MA 02138, USA}

\author{N.~van~der~Marel}
\affiliation{Institute for Astronomy, University of Hawaii at Manoa, Honolulu, HI, USA}

\author{E.~F.~van Dishoeck}
\affiliation{Leiden Observatory, Leiden University, P.O. Box 9513, 2300RA Leiden, The Netherlands}
\affiliation{Max-Plank-Institut f\"{u}r Extraterrestrische Physik, Giessenbachstra\ss e 1, D-85748 Garching, Germany}

\author{S.~Ataiee}
\affiliation{Center for Space and Habitability, Physikalisches Institut, Universitaet Bern, 3012 Bern, Switzerland}

\author{M.~Benisty}
\affiliation{Univ. Grenoble Alpes, CNRS, IPAG, F-38000 Grenoble, France}

\author{T.~Birnstiel}
\affiliation{University Observatory, Faculty of Physics, Ludwig-Maximilians-Universit\"{a}t M\"{u}nchen, Scheinerstr. 1, 81679 Munich, Germany}

\author{A.~Juh\'asz}
\affiliation{Institute of Astronomy, Madingley Road, Cambridge CB3 OHA, UK}

\author{A.~Natta}
\affiliation{Dublin Institute for Advanced Studies, School of Cosmic Physics, 31 Fitzwilliam Place, Dublin 2, Ireland}
\affiliation{INAF-Arcetri, Largo E. Fermi 5, I-50125 Firenze}

\author{L.~Ricci}
\affiliation{Harvard-Smithsonian Center for Astrophysics, 60 Garden Street, Cambridge, MA 02138, USA}
\affiliation{Department of Physics and Astronomy, Rice University, 6100 Main Street, 77005 Houston, TX, USA}

\author{L.~Testi}
\affiliation{INAF-Arcetri, Largo E. Fermi 5, I-50125 Firenze}
\affiliation{European Southern Observatory, Karl-Schwarzschild-Str. 2, D85748 Garching, Germany}

\begin{abstract}
We present new Atacama Large Millimeter/sub-millimeter Array (ALMA) 1.3\,mm continuum observations of the SR\,24S transition disk with an angular resolution $\lesssim0.18''$ (12\,au radius).  We perform a multi-wavelength investigation by combining new data with previous ALMA data at 0.45\,mm.  The visibilities and images of the continuum emission at the two wavelengths are well characterized by a ring-like emission. Visibility modeling finds that the ring-like emission is narrower at longer wavelengths, in good agreement with models of dust trapping in pressure bumps, although there are complex residuals that suggest potentially asymmetric structures. The 0.45\,mm emission has a shallower profile inside the central cavity than the 1.3\,mm emission.  In addition, we find that the $^{13}$CO and C$^{18}$O (J=2-1) emission peaks at the center of the continuum cavity.  We do not detect either continuum or gas emission from the northern companion to this system (SR\,24N), which is itself a binary system. The upper limit for the dust disk mass of SR\,24N is  $\lesssim 0.12\,M_{\bigoplus}$, which gives a disk mass ratio in dust between the two components of  $M_{\mathrm{dust, SR\,24S}}/M_{\mathrm{dust, SR\,24N}}\gtrsim840$. The current ALMA observations may imply that either planets have already formed in the SR\,24N disk or that dust growth to mm-sizes is inhibited there and that only warm gas, as seen by ro-vibrational CO emission inside the truncation radii of the binary, is present.

\end{abstract}


\section{Introduction}     \label{introduction}

Recent multi-wavelength observations of protoplanetary disks revealed astonishing structures, such as concentric dust rings, spiral arms, and asymmetries \citep[e.g.][]{marel2013, alma2015, casassus2015, andrews2016, stolker2016, deboer2016, ginski2016, perez_L2016, fedele2017}. These observations suggest that significant evolution has taken place and that probably planets have already imprinted their existence in the parental disks. Transition disks (TD) have been of particular interest due to their inner cavities, which were first identified by the lack of infrared emission \citep{strom1989}. Different mechanisms have been proposed for the origin of TD cavities, including photo-evaporation, magneto-rotational instabilities (MRI), and planet-disk interaction \citep[e.g.][]{regaly2012, zhu2012, alexander2014, flock2015, dipierro2016, pinilla2016b}. To understand whether one or several mechanisms dominate the evolution, it is crucial to spatially resolve protoplanetary disks at different wavelengths since each physical process (or the combination of several of them) lead to different structures for the small/large dust and for the gas \citep[e.g.][]{rosotti2013}.

For instance, when a planet opens a gap in a gaseous disk, at the outer edge of the gap, the gas density increases and the pressure has a local maximum where dust particles stop their fast radial drift and accumulate \citep[e.g.][]{whipple1972, pinilla2012}. This process can lead to a spatial difference for the distribution of small (micron-sized) particles, which are well coupled to the gas, and large (mm-sized) particles. As a consequence, smaller and less depleted cavities or gaps are expected in the gas and small grains than in the large mm-dust particles. In this scenario, the possibility of observing rings and cavities in the dust at different wavelengths strongly depends on the disk viscosity \citep{maria2016}. Similarly, the formation of a broad and robust pressure bump that can arise from MRI processes, such as dead zones together with MHD winds, can lead to  comparable structures in the gas and dust as in the case of planet-disk interaction \citep{pinilla2016b}. 

Recent ALMA gas and dust observations of TDs show that in most cases, there is gas inside the mm-dust cavity. The gas usually also shows a cavity, but with a lower depletion factor than the mm-emission \citep[e.g.][]{bruderer2014, perez_s2015, canovas2016, marel2015, marel2016}. In this paper, we report ALMA  observations of the transition disk around SR\,24S at 1.3\,mm of the dust continuum emission and the molecular lines $^{13}$CO (J=2-1) and  C$^{18}$O (J=2-1). For the analysis, we combine these new data with previous ALMA  data at 0.45\,mm.

\begin{figure*}
 \centering
   	\includegraphics[width=15cm]{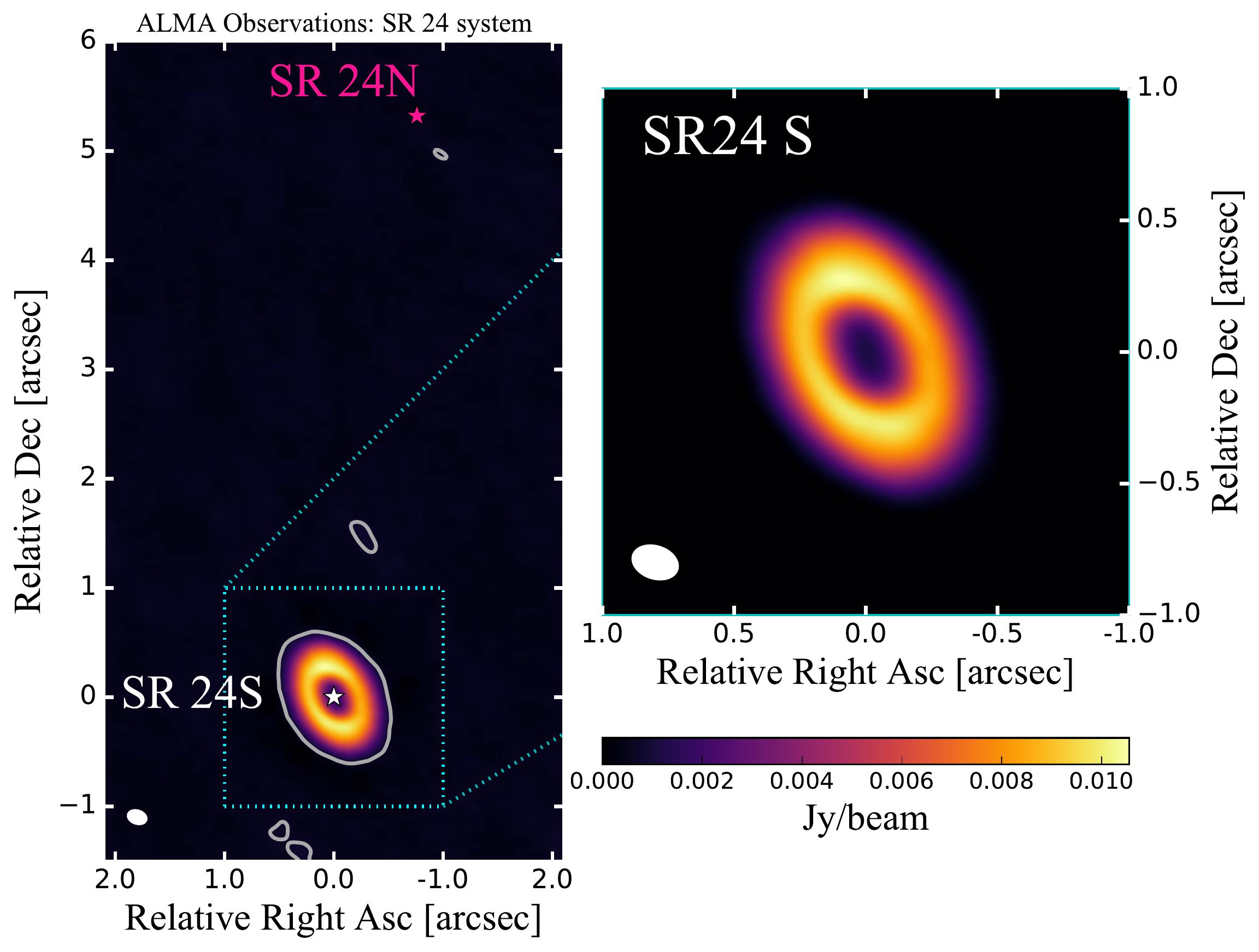}   
   \caption{ALMA dust continuum observations of the SR\,24 system at 1.3\,mm with a resolution of 0.18''$\times$0.12''. The contour lines in the left panel correspond to $3\sigma_N$, where $\sigma_N=75\,\mu\rm{Jy\,beam}^{-1}$ is the rms measured at the location of the north component. The right panel corresponds to a zoom-in of the image centered at the location of the south component SR\,24S.}
   \label{figure0}
\end{figure*}

SR\,24 is a hierarchical triple system located in the L1688 dark cloud in the Ophiuchus star-formation region. L1688  extends over a range of distances, likely between 120\,pc and 145\,pc \citep{mamajek2008, loinard2008}. In this paper, we adopt a value of 137\,pc. Each of the components of SR\,24 was identified as a T-Tauri star \citep{wilking1989, ghez1993}, with infrared excess \citep{greene1994}. The separation between the two main components of SR\,24 is 5.2'' at a position angle (PA) of 348$^{\circ}$ \citep{reipurth1993}.  The northern component, SR\,24N, is itself a binary system with a separation of 0.2'' at a PA of 87$^{\circ}$ \citep{simon1995}.  The primary component SR\,24S is a K2 star and its mass is $>1.4\,M_{\odot}$, while the stars in SR\,24N are a K4-M4 star with a mass of $0.61^{+0.6}_{-0.27}\,M_\odot$ and a K7-M5 star with a mass of $0.34^{+0.46}_{-0.18}\,M_\odot$ \citep{correia2006}. 

SR\,24S and SR\,24N show similar infrared emission, indicating warm dust in the inner part of both disks \citep{stanke2000, bontemps2001}. \cite{brown2013} reported ro-vibrational CO emission  at 4.7\,$\mu$m tracing warm gas in the inner parts of both disks (SR\,24S and SR\,24N). In addition, both circumprimary (SR\,24S) and circumsecondary  (SR\,24N) disks were resolved in the infrared image obtained with the adaptive optics coronagraph CIAO (in the Subaru telescope). These observations show that the primary disk is more extended than the secondary and the disks seem to be extended enough to fill the effective Roche radius of the system  \citep{mayama2010}. SR\,24S and SR\,24N are highly accreting, the accretion rates obtained from the hydrogen recombination lines are for SR\,24N $10^{-6.9}\,M_\odot$year$^{-1}$ and for  SR\,24S $10^{-7.15}\,M_\odot$year$^{-1}$ \citep{natta2006}.

However, only the southern component, SR\,24S, has been detected in the dust continuum \citep{nuernberger1998}. \cite{andrews2005} reported SMA observations of SR\,24 at 225\,GHz (1.3\,mm) continuum and $^{12}$CO (J=2-1) line emission. Stronger CO emission was seen around SR\,24N than around SR\,24S.  Later SMA observations of SR\,24S at 0.88\,mm detected a dust cavity of $\sim32$\,au radius \citep{andrews2010}, revealing that SR\,24S is a transition disk.  \cite{marel2015} reported ALMA Cycle 0 observations of the continuum and $^{12}$CO (J=6-5) at 0.45\,mm. The CO observations of SR\,24S were affected by extended emission and foreground absorption from the dark cloud in Ophiuchus, and it was not possible to infer the amount of gas and its distribution inside the dust-cavity. The dust cavity size was inferred to be 25\,au from the fitting of the ALMA Cycle~0 continuum observations. 

This paper is organized as follows. In Sect.~\ref{observations} and Sect.~\ref{results}, we summarize the details of the ALMA observations, data reduction and imaging. Sect.~\ref{analysis} presents the analysis of the data, in particular of the continuum emission and the comparison with previous ALMA observations at 0.45\,mm. The discussion and main conclusions are in Sect.~\ref{discussion} and Sect.~\ref{conclusion} respectively. 

\section{Observations}     \label{observations}
SR\,24 was observed with ALMA in Band 6 during Cycle 2 on September 26th, 2015 (\#2013.1.00091.S).  For these observations 34 antennas were used and the longest baseline was 2269.9\,m. The source was observed in four spectral windows, each with a bandwidth of 1875.0\,MHz. Two of them were chosen with a smoothed resolution of 976.563\,kHz centered at 219.56035\,GHz for the C$^{18}$O (J=2-1) transition  and 220.39868\,GHz for the $^{13}$CO (J=2-1) transition, for a channel width of $\sim1.35$\,km\,s$^{-1}$ for the two lines. The other spectral windows were configured to obtain the continuum emission centered at 235\,GHz ($\sim$1.3\,mm). The quasar QSO J1517-2422 was observed for bandpass calibration, while the quasars QSO J1617-2537 and QSO J1627-2426  were observed for phase calibration. The asteroid Pallas was observed for the flux calibration. The total observation time was 49.31\,min, with a total time on source of 22.67\,min. The data were calibrated using the Common Astronomy Software Package (CASA), version 4.4. For the reduction, there was one antenna flagged due to strange/elevated $T_{\rm{sys}}$.

\begin{figure*}
 \centering
 \tabcolsep=0.1cm 
   \begin{tabular}{ccc}   
   	\includegraphics[width=6cm]{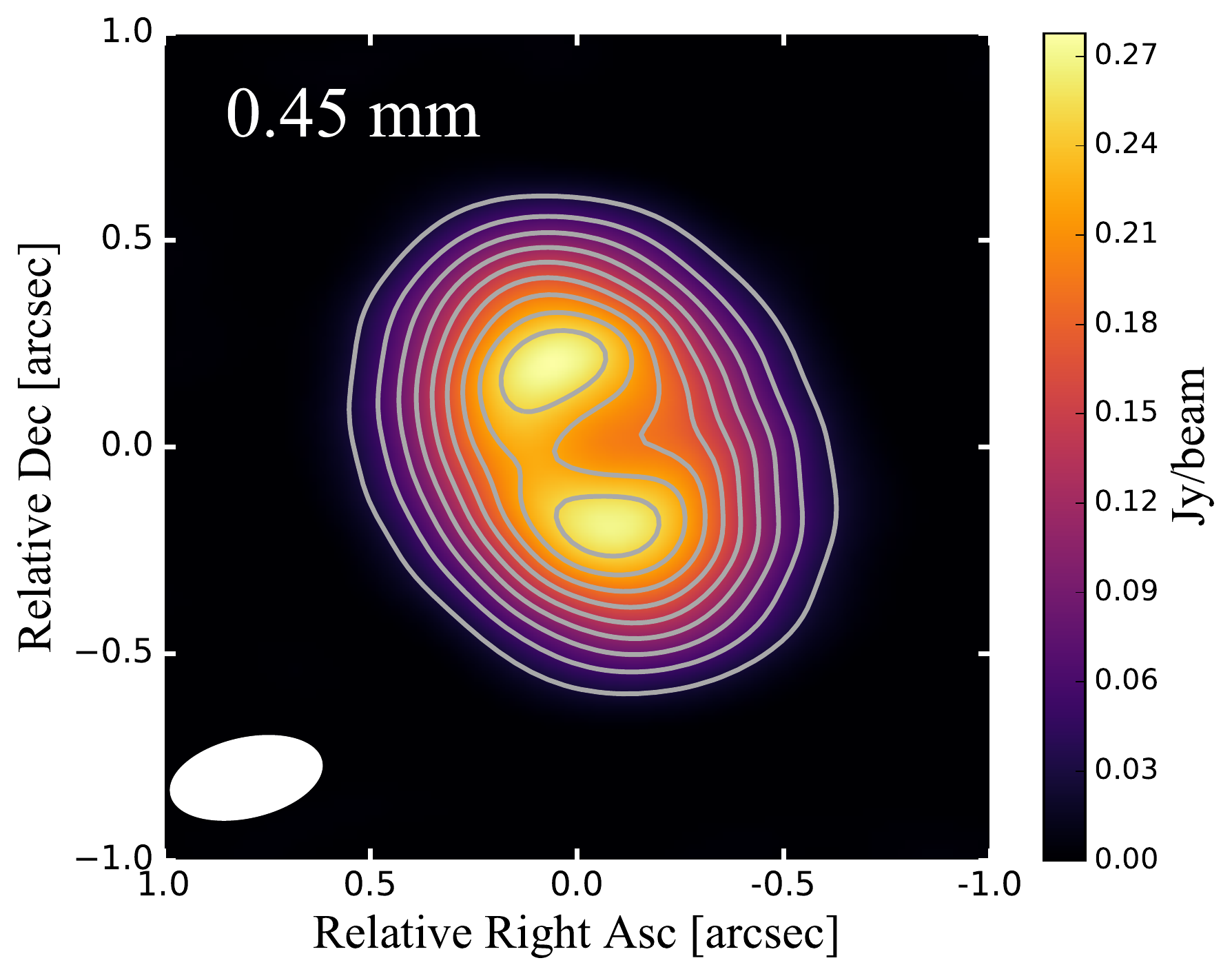}&\includegraphics[width=6cm]{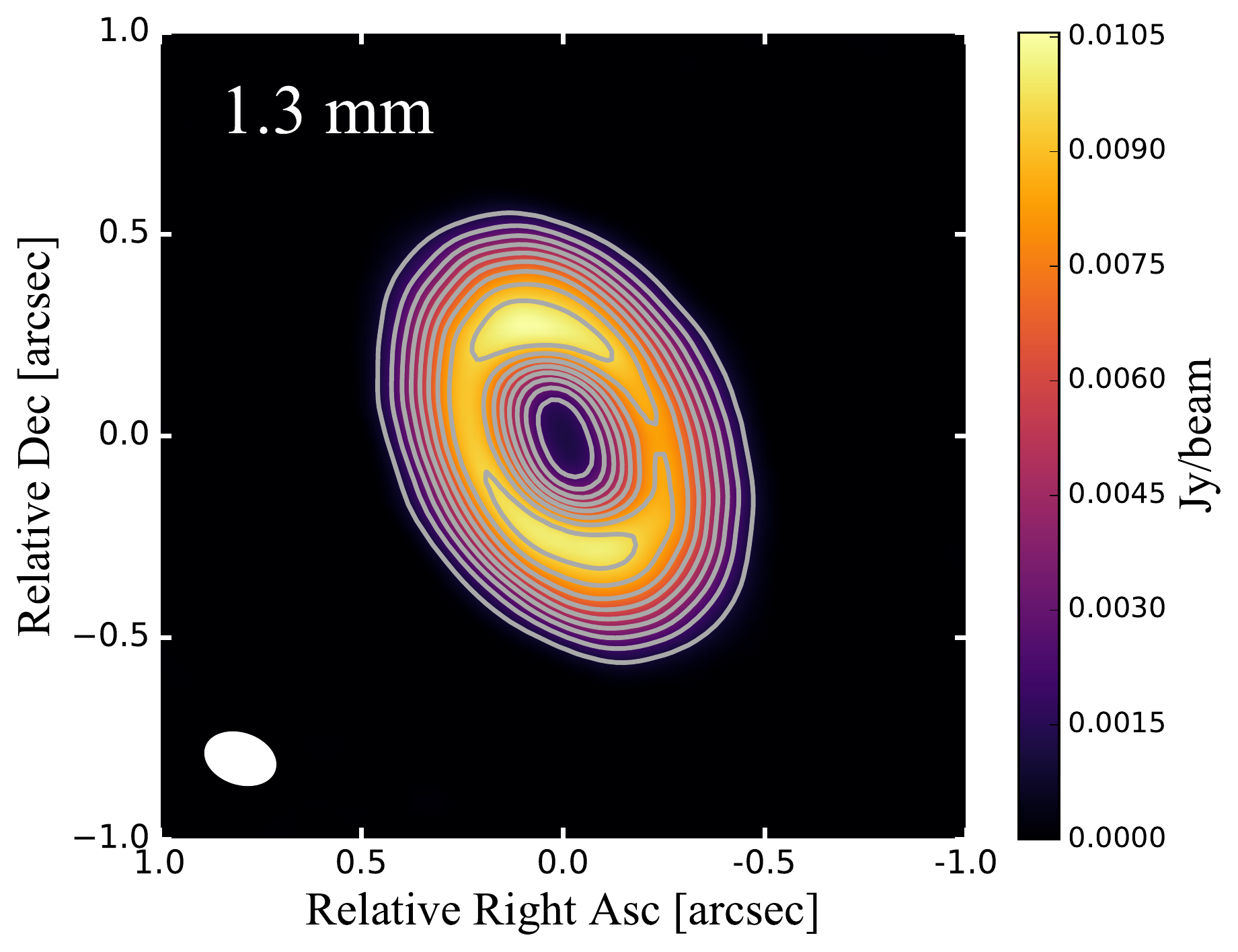}&\includegraphics[width=6cm]{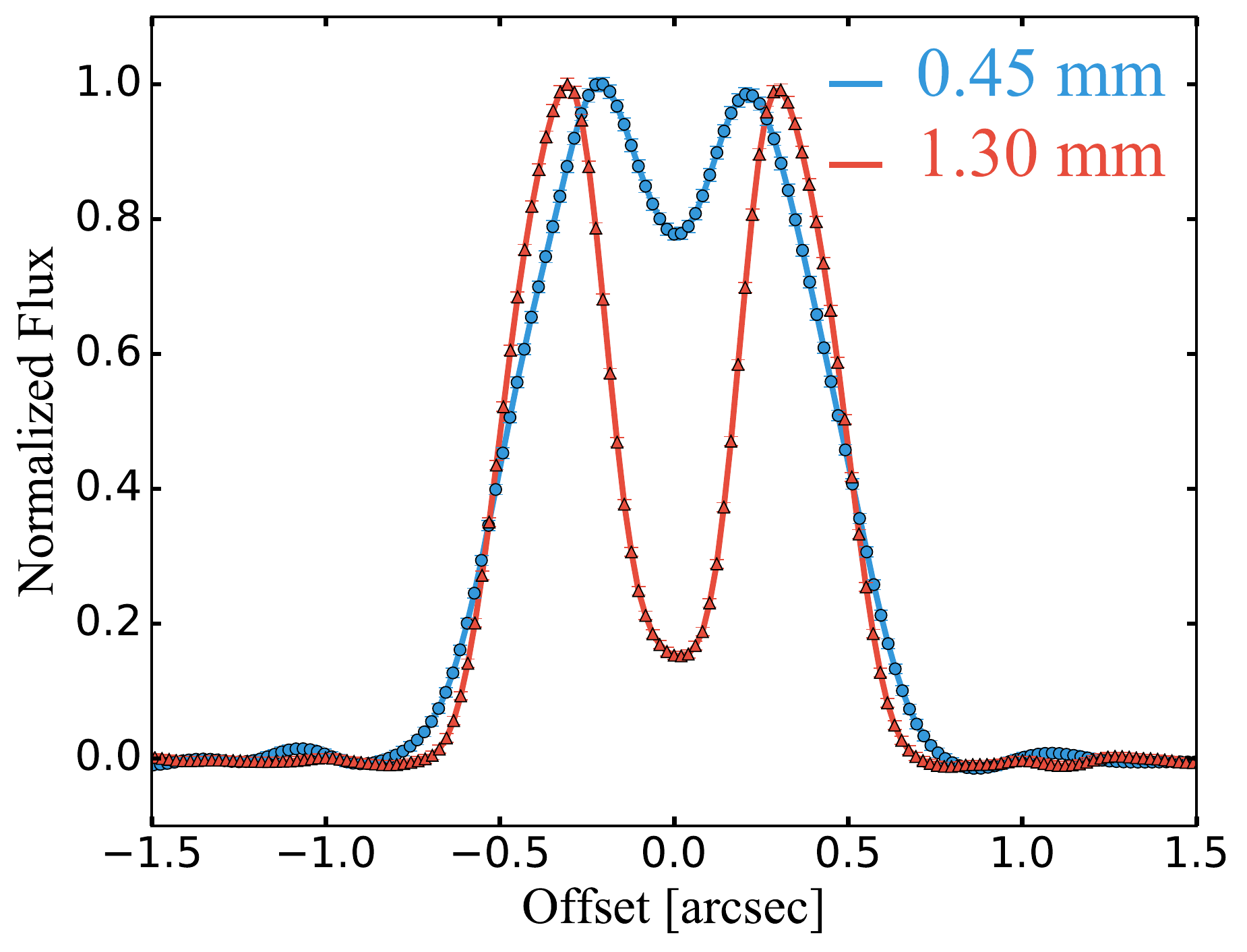}
   \end{tabular}
    \caption{Continuum observations of the transition disk SR\,24S. \emph{Left panel: }ALMA continuum map at 0.45\,mm \emph{Middle panel:} ALMA  continuum map at 1.3\,mm. Specific details are summarized in Table~\ref{tbl:obsproperties}. For both maps, the contour  levels are 10, 20, . . . , 90\% of the peak of emission. \emph{Right panel: } Normalized continuum flux at 0.45 and 1.3\,mm along the PA=24$^\circ$ of the disk (error bars are also included, which are of the size of the points).}
   \label{continuum_observations}
\end{figure*}

For imaging, the data were correctly centered by two independent procedures. First, to find the center of the image, position angle (PA), and disk inclination ($i$), a simple Gaussian or disk model was used to fit the data (using {\tt uvmodelfit} in CASA), either using only short baselines ($\lesssim200\,$k$\lambda$) or all the uv coverage. The obtained PA and inclination for both models of the disk are $27.8^{\circ}\pm{1.3}^{\circ}$ and $49.8^{\circ}\pm{2.4}^{\circ}$ respectively. We applied the same procedure for fitting the 0.45\,mm data with {\tt uvmodelfit}, finding PA=$26.9^{\circ}\pm{1.6}^{\circ}$ and $i=47.2^{\circ}\pm{3.1}^{\circ}$, in agreement with the values found by \cite{marel2015}. Second, since the image does not have significant asymmetries, the center is also checked by minimizing the rms scatter of  the imaginary part of the visibilities around zero. Both methods give very similar centers, and $\alpha_{2000}$=16:26:58.5, $\delta_{2000}$=-24:45:37.2 were used to correct the phase center and obtain the visibilities using {\tt fixvis}. The same procedure was used for previous ALMA observations at 0.45\,mm. Nevertheless, the center, PA and $i$ are again taken as free parameters when the analysis is done in the visibility domain (Sect.~\ref{analysis}). Different studies show that visibility modeling has some advantages over imaging analysis, since it can identify unresolved structures and constrain better the disk morphology, without being limited by deconvolution issues that may arise during imaging \citep[e.g.][]{perez_L2014, walsh2016, white2016, zhang2016}.

Continuum and line imaging were performed using the {\tt clean}  algorithm. We used natural weighting and Briggs weighting (robust = 0.5) to find the best compromise between resolution and sensitivity. For the continuum, with Briggs weighting, we achieved a rms of 63\,$\mu$Jy\,beam$^{-1}$ with a beam size of 0.18'' $\times$ 0.12''.  The continuum was subtracted from line-containing channels using {\tt uvcontsub}. Since the $^{13}$CO and the C$^{18}$O are weak detections, we performed natural weighting for a final rms of around 1.4\,mJy\,beam$^{-1}$ per 1.35\,km\,s$^{-1}$ channel for both lines, and the final beam size in this case is 0.21''$\times$0.16''.\\ \\

\section{Results}     \label{results}

\begin{table}
\caption{Properties of the continuum ALMA images of SR\,24S (Fig.~\ref{continuum_observations})}
\label{tbl:obsproperties}
\centering            
\begin{tabular}{c|| c | c | c | c | c  }
\hline
\hline
{\footnotesize Cycle}&{\footnotesize beam}&{\footnotesize $\lambda$}&{\footnotesize $F_{\rm{peak}}$}&{\footnotesize $F_{\rm{total}}$}&{\footnotesize $\sigma$}\\
&{\footnotesize ('')}&{\footnotesize (mm)}&{\footnotesize (mJy)}&{\footnotesize (mJy)}&{\footnotesize (mJy\,beam$^{-1}$)}\\
\hline
{\footnotesize 0}&{\footnotesize 0.37$\times$0.19}&{\footnotesize 0.45}&{\footnotesize 278}&{\footnotesize 1885}&{\footnotesize1.9}\\
\hline
{\footnotesize 2}&{\footnotesize 0.18$\times$0.12}&{\footnotesize 1.30}&{\footnotesize 15.4}&{\footnotesize 220}&{\footnotesize0.06}\\
\hline
\end{tabular}
\end{table}

\subsection{Continuum emission} \label{sect:continuum}
Figure~\ref{figure0} show the resulting 1.3\,mm image of the SR\,24 system after the cleaning process (and after primary beam correction). In the left panel, the contour lines correspond to $3\sigma_N$, where $\sigma_N=75\,\mu\rm{Jy\,beam}^{-1}$ is the rms measured at the location of the north component \citep[around 16:26:58.44 -24:45:31.9, ][]{cutri2003}. Taking a circular area with a radius of 1'' and centered at the position of SR\,24N, the total flux is $\sim 3.0\sigma_N$. Assuming optically thin emission, the dust mass can be estimated as \citep{hildebrand1983}

\begin{equation}
	M_{\mathrm{dust}}\simeq\frac{{d^2 F_\nu}}{\kappa_\nu B_\nu (T(r))}
  \label{mm_dust_mass}
\end{equation}

\noindent where $d$ is the distance to the source, $\kappa_\nu$ is the dust opacity at a given frequency, and $B_\nu (T)$ is the Planck function for a given temperature radial profile $T(r)$. Assuming a distance of 137\,pc, a dust opacity at 1.3\,mm of $\sim3\,$cm$^{2}$g$^{-1}$ \citep[e.g.][]{andrews2011} and a temperature of 20\,K, the upper limit for the dust mass of the SR\,24N disk is  $M_{\mathrm{dust, SR\,24N}}\lesssim 3.5\times10^{-7}\,M_\odot$ or in Earth masses equivalent to  $\lesssim 0.12\,M_{\bigoplus}$.

\begin{figure*}
\centering
 \tabcolsep=0.1cm 
   \begin{tabular}{cc}   
   \includegraphics[width=13cm]{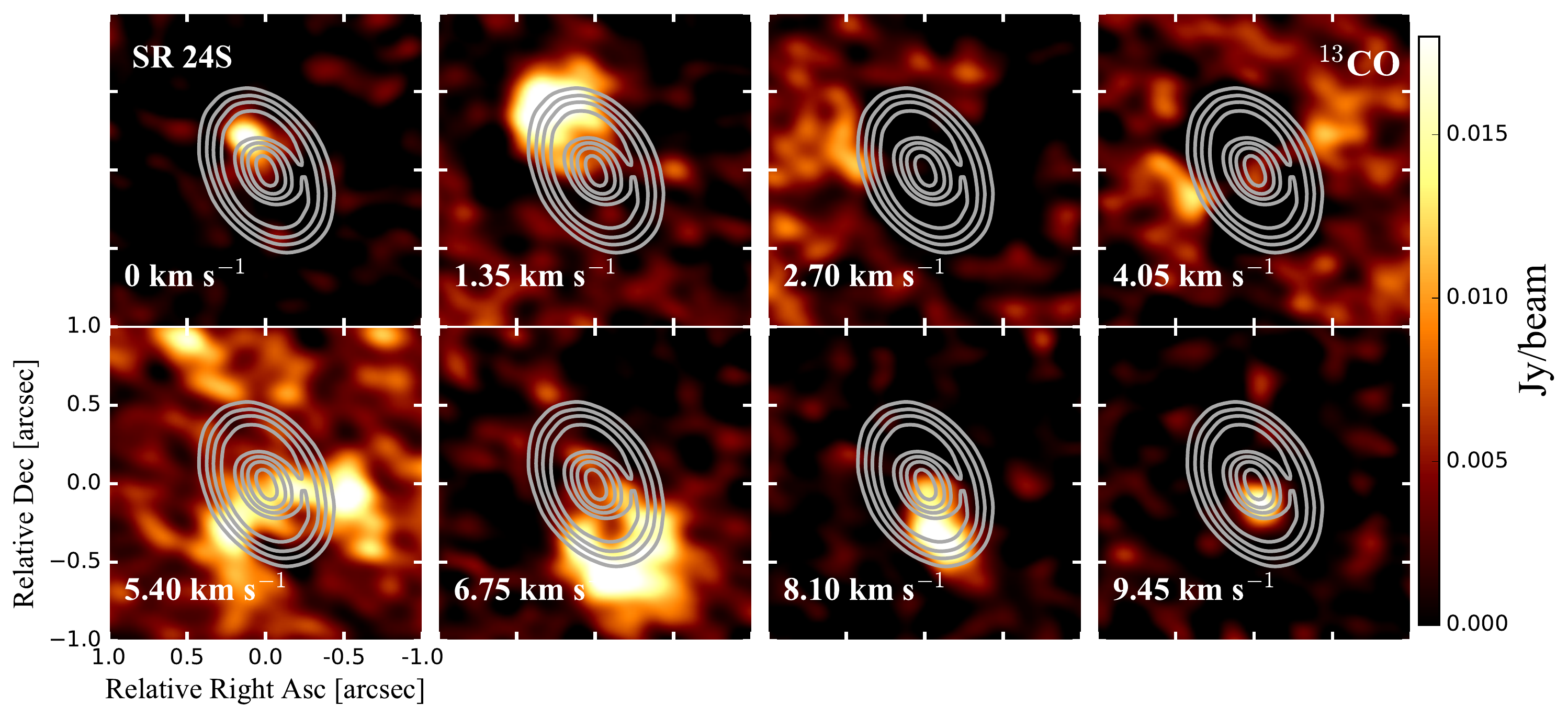}&\includegraphics[width=5cm]{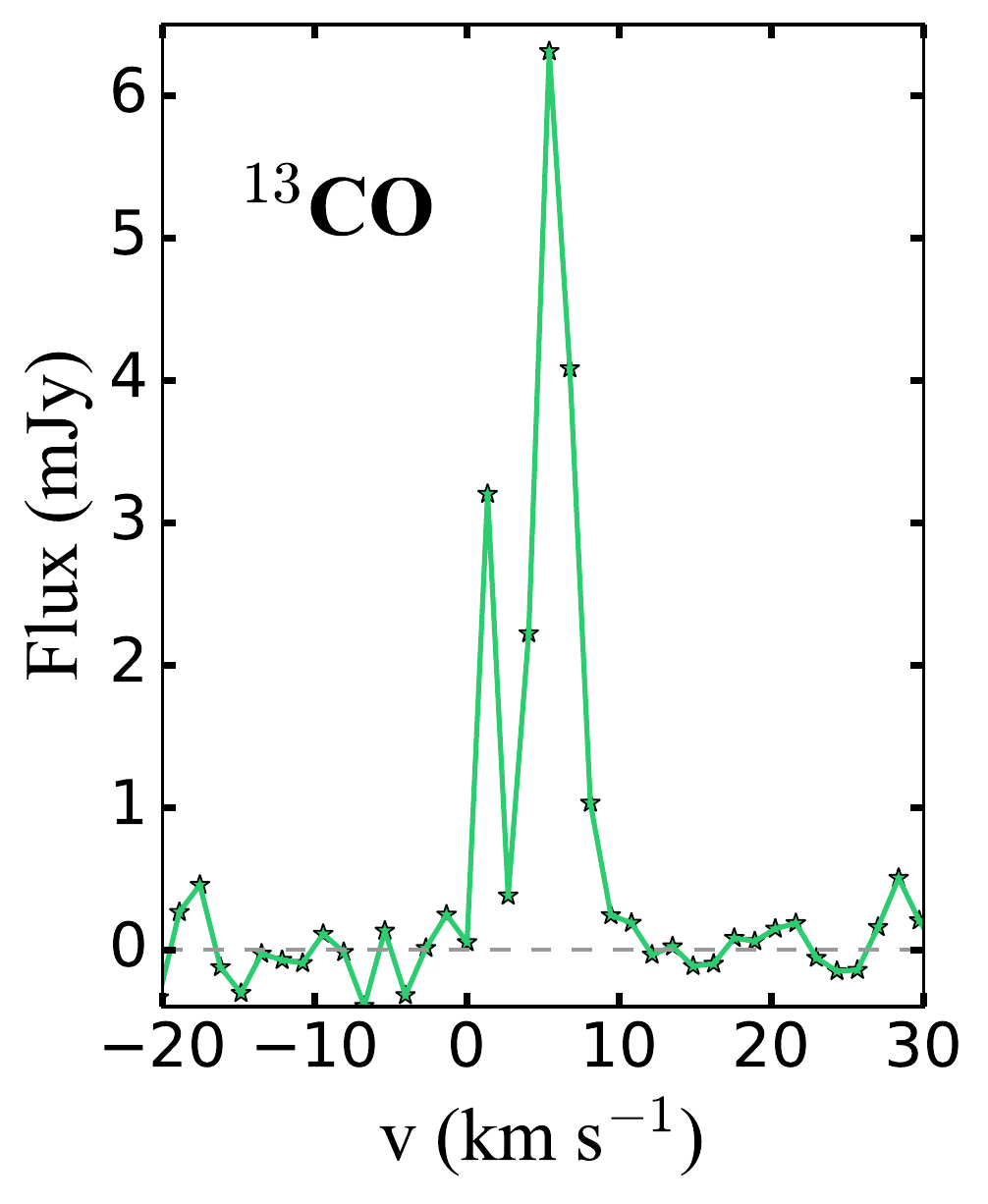}\\
   \includegraphics[width=13cm]{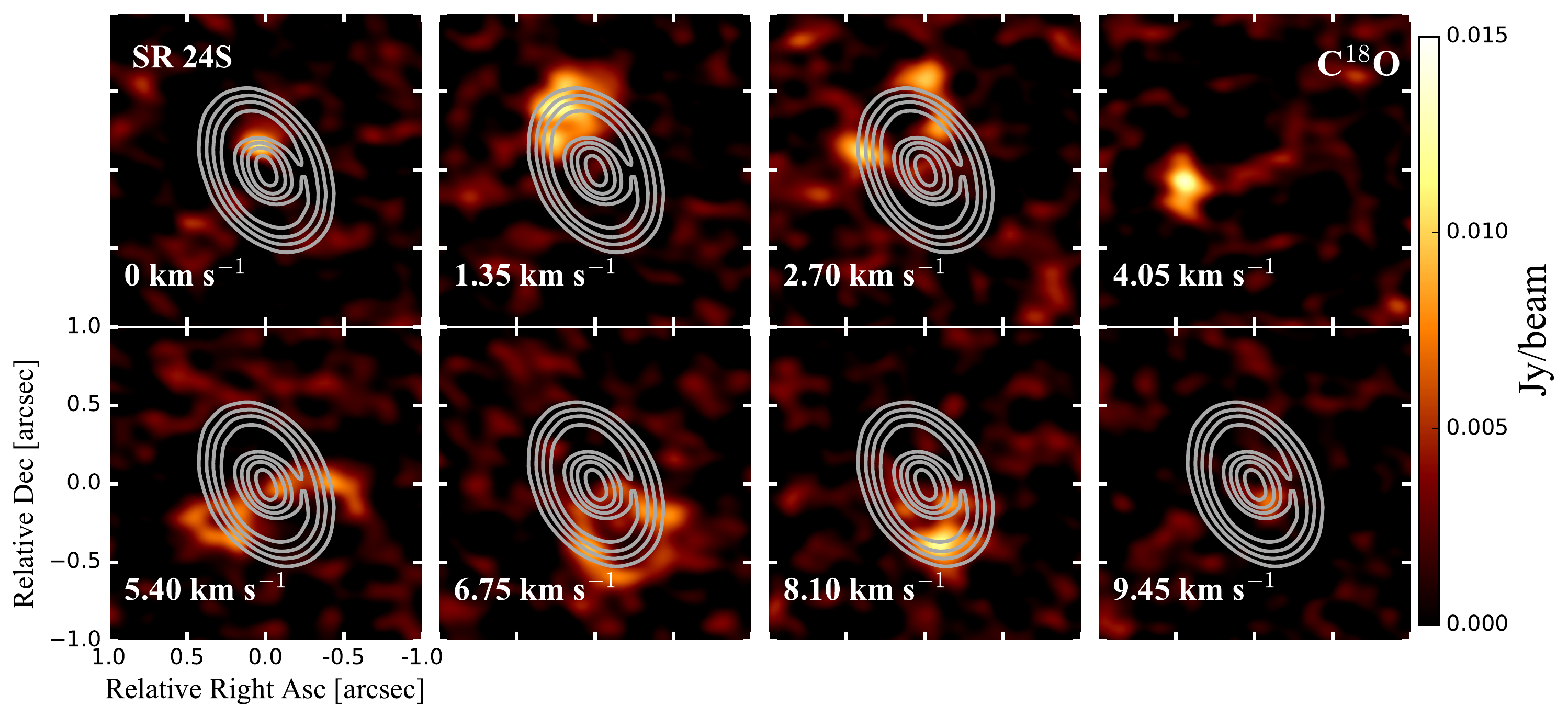}&\includegraphics[width=5cm]{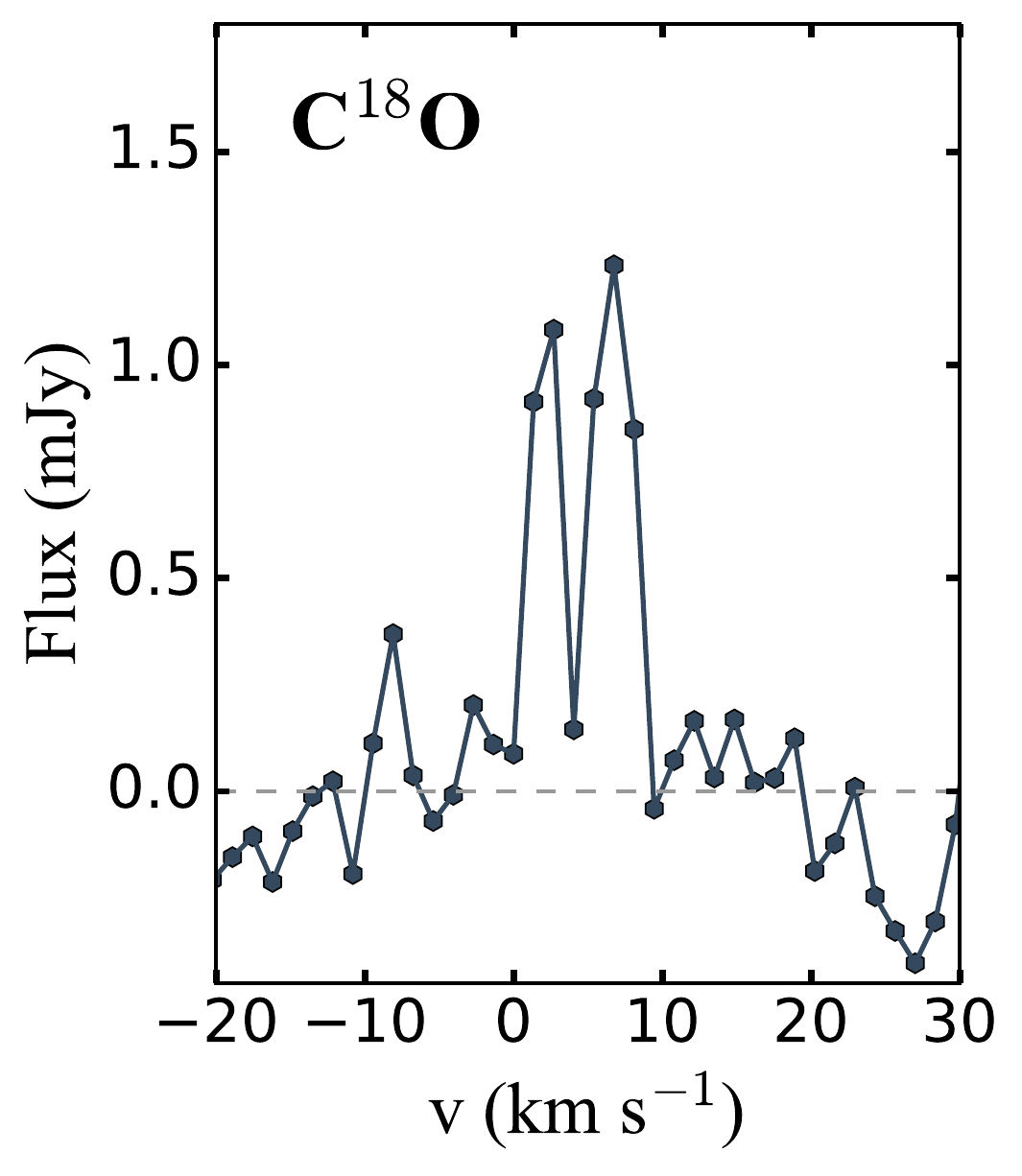}
   \end{tabular}
    \caption{\emph{Left panels:} Channel maps of the $^{13}$CO (\emph{top panels}) and C$^{18}$O (\emph{bottom panels}) emission in SR\,24S from 0 to 9.45\,km\,s$^{-1}$. The spectral resolution is 1.35\,km\,s$^{-1}$ and the rms  is around 1.4\,mJy\,beam$^{-1}$ per channel for both lines. The contour  levels are 20, . . . , 100\% of the peak of the dust continuum emission. \emph{Right panels:} $^{13}$CO and C$^{18}$O (J=2-1) spectrum by integrating over a circular area centered at the location of SR\,24S and with a radius of 1''.}
   \label{channel_maps}
\end{figure*}

Figure~\ref{continuum_observations} presents the continuum ALMA observations of SR\,24S at 0.45\,mm and at 1.3\,mm.  The details of the calibration process for the Cycle~0 data are presented in \cite{perez_L2014}. We summarize the properties of each image in Table~\ref{tbl:obsproperties}. The continuum emission is detected with a signal-to-noise ratio (with respect to the peak) of 146  for 0.45\,mm data  and 256 for the 1.3\,mm data (see Table~\ref{tbl:obsproperties}). The total flux at 0.45\,mm is 1.9\,Jy and at 1.3\,mm 0.22\,Jy.  Calculating the dust disk mass from 1.3\,mm flux, and assuming the same opacity and temperature as for SR\,24N, we obtain that $M_{\mathrm{dust, SR\,24S}}\sim3.0\times10^{-4}M_\odot$, implying a  dust disk ratio between the southern and the northern component of $\gtrsim840$. Nonetheless, Eq.~\ref{mm_dust_mass} assumes optically thin emission, which may not be the case for SR\,24S, specially close to the location of the ring (see. Sect.~\ref{sect:spectral_index}). If only part of the emission is optically thin, the dust mass for the disk around SR\,24S is underestimated, which would increase the dust mass disk ratio between the southern and the northern component.

Taking the total flux at each wavelength, the integrated spectral index is given by\\  $\alpha_{\rm{mm}} = \ln(F_{\rm{1.3\,mm}}/F_{\rm{0.45\,mm}}) /\ln (\rm{0.45\,mm}/\rm{1.3\,mm}) = 2.02\pm{0.13}$ (the error includes a calibration uncertainty of 10\%), which is lower than the value  previously reported \citep{pinilla2014} based on SMA and ATCA observations at 0.88\, and 3.0\,mm \citep{andrews2011, ricci2010}. This low value may indicate grain growth and/or a small cavity, but most likely arises from optically thick emission as discussed in Sect~\ref{sect:spectral_index}. 

Figure~\ref{continuum_observations} also shows the continuum flux normalized to the peak of emission at 0.45 and 1.3\,mm of a radial cut along the PA of the disk (obtained in Sect.~\ref{analysis}). Both profiles reveal a cavity and a ring. The 1.3\,mm emission strongly decreases inside the cavity where the flux is reduced by around 85\%. Contrary, the 0.45\,mm emission shows a shallower cavity with the emission reduced by about 24\% compared to the peak of emission. In addition, the position of the peak of the ring is located further out at 1.3\,mm. However, this contrast and location of the cavity can be affected by the beam convolution and a more detailed analysis of the intensity profiles is done in the visibility domain in Sect.\ref{sect:morphology}.

\begin{figure*}
 \centering
 \tabcolsep=0.1cm 
   \begin{tabular}{ccc}   
   	\includegraphics[width=6cm]{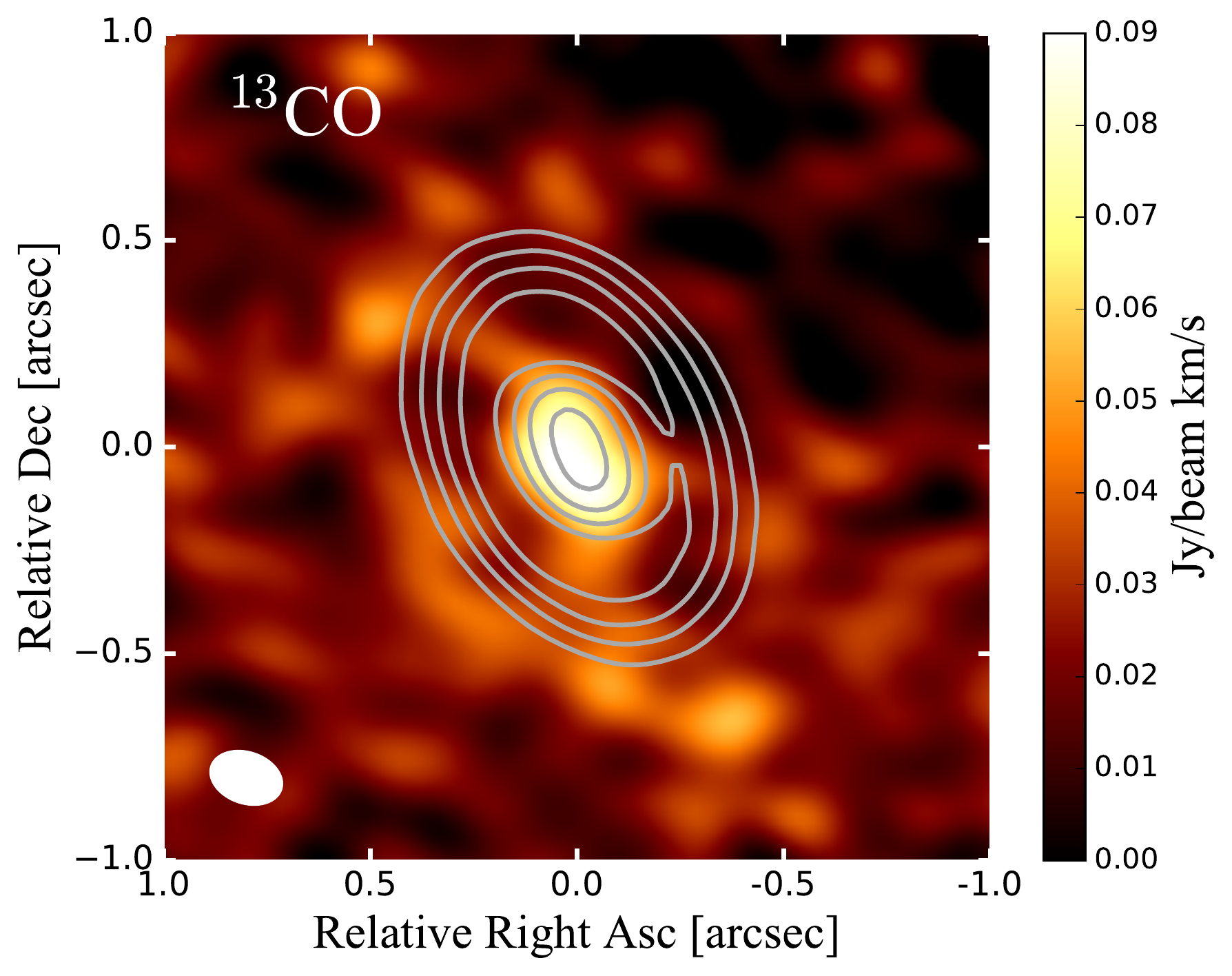}&\includegraphics[width=6cm]{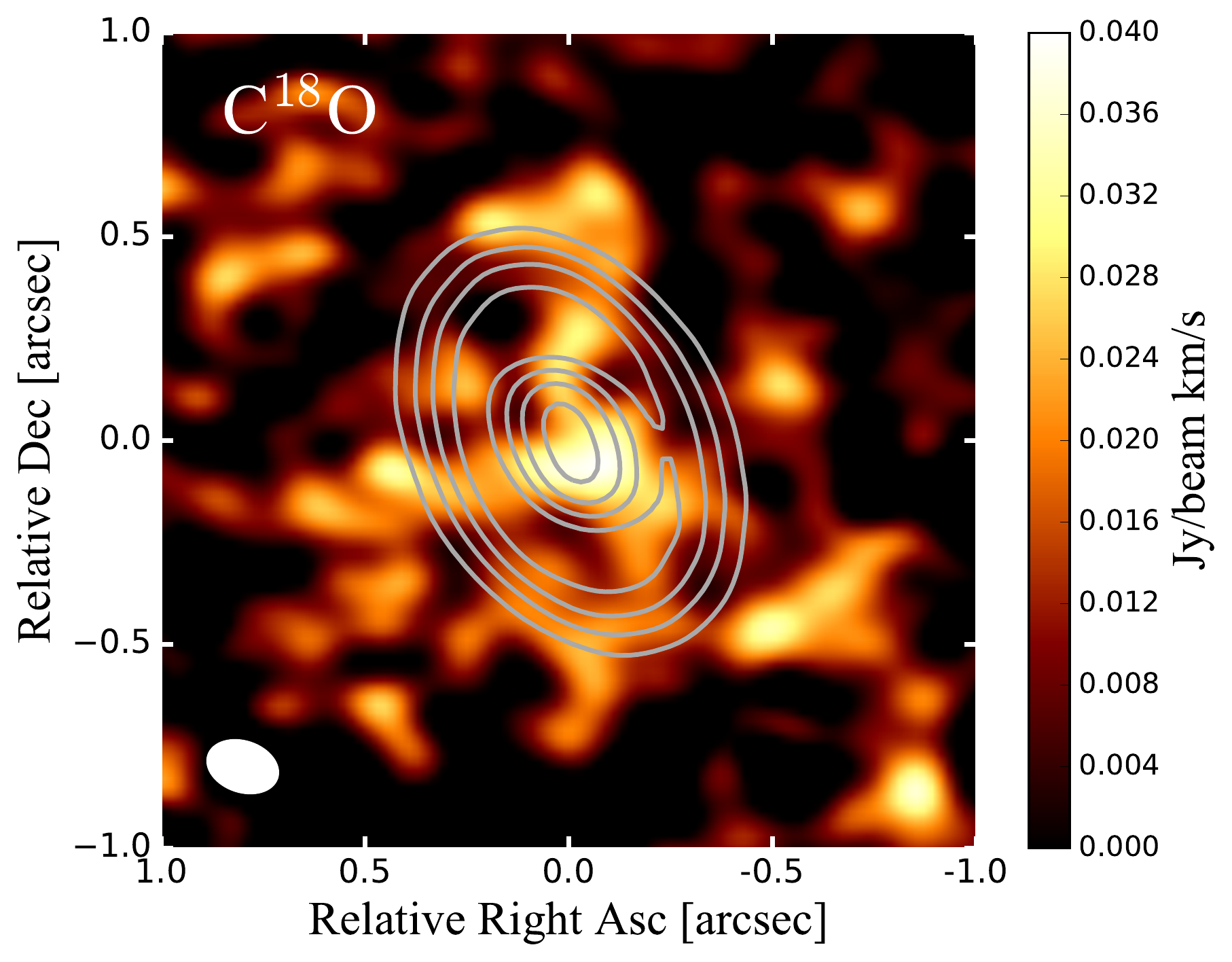}&\includegraphics[width=6cm]{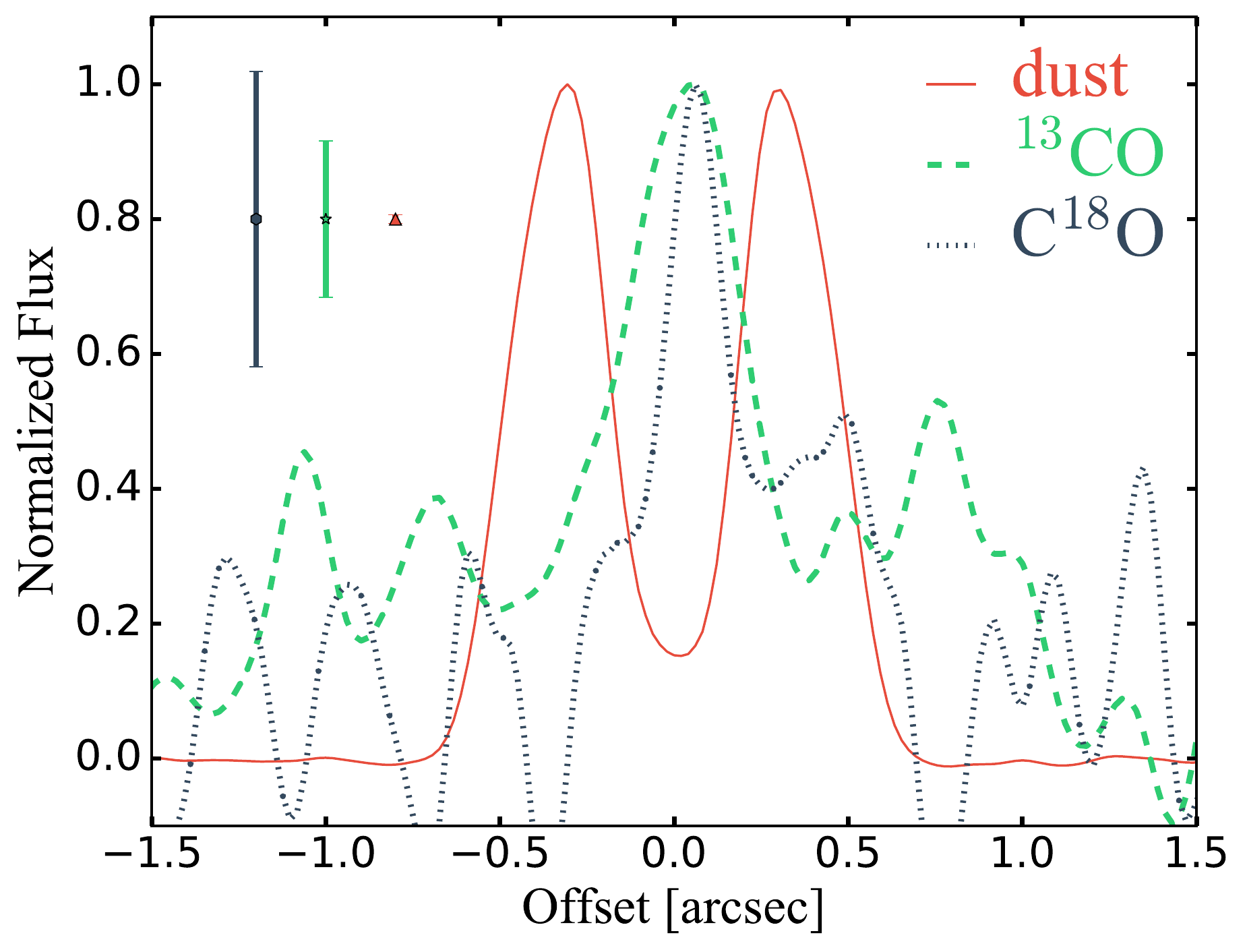}
   \end{tabular}
    \caption{Zero moment map of  $^{13}$CO \emph{(left panel)} and C$^{18}$O \emph{(middle panel)} of the transition disk SR\,24S. The contour  levels are 20, . . . , 100\% of the peak of the dust continuum emission. \emph{Right panel: } Normalized flux along the PA of the disk  from both zero moment maps. For comparison the flux of the continuum emission is over-plotted.}
   \label{continuum_vs_gas}
\end{figure*}

\subsection{Gas emission}
Both $^{13}$CO and C$^{18}$O (J=2-1) lines are detected, but affected by foreground absorption from the nearby dark cloud. In particular $^{13}$CO is more contaminated than C$^{18}$O, because C$^{18}$O is more optically thin than $^{13}$CO. Figure~\ref{channel_maps} shows the channel maps of the $^{13}$CO and C$^{18}$O emission in SR\,24S from 0 to 9.45\,km\,s$^{-1}$. In addition, the $^{13}$CO and C$^{18}$O (J=2-1) spectrum is shown in Fig.~\ref{channel_maps}, which is obtained by integrating over a circular area centered at the location of SR\,24S and with a radius of 1''. These channel maps confirm the presence of $^{13}$CO and C$^{18}$O in the SR\,24S disk, but also the effect of the foreground absorption, in particular in the channels of 2.7 and 4.05 \,km\,s$^{-1}$. Thus, the asymmetry of the double-peaked velocity profile for both lines is likely due to this foreground absorption. 

Figure~\ref{continuum_vs_gas} shows the zero moment map for $^{13}$CO and C$^{18}$O obtained from -1.35\,km\,s$^{-1}$ to 10.8\,km\,s$^{-1}$, where the channels contain significant emission ($\gtrsim 5\,\sigma$, with $\sigma=1.4\,$mJy\,beam$^{-1}$ per 1.35\,km\,s$^{-1}$ channel). For $^{13}$CO,  inside a circle of $\sim0.7''$ radius centered at the location of SR\,24S,  the total flux per velocity interval is $\sim1\,$Jy\,beam$^{-1}$\,km\,s$^{-1}$, and approximately  60\% of this emission comes from the inner part of the disk inside the mm-dust cavity. The emission peaks at the center with a value of $\sim0.1\,$Jy\,beam$^{-1}$\,km\,s$^{-1}$ and the rms of the zero moment map is around $\sim10\,$mJy\,beam$^{-1}$\,km\,s$^{-1}$, which gives a signal-to-noise ratio of 10 with respect to the peak and 100 with respect to the total flux. For C$^{18}$O, inside the same circle of $\sim0.7''$ radius, the total flux per velocity is $\sim0.5\,$Jy\,beam$^{-1}$\,km\,s$^{-1}$, and approximately  40\% of this emission comes from the inner part. This emission also peaks at the center with a value of $\sim40\,$mJy\,beam$^{-1}$\,km\,s$^{-1}$ and the rms of the zero moment map is around $\sim8.6\,$mJy\,beam$^{-1}$\,km\,s$^{-1}$, which gives a signal-to-noise ratio of 5 and 58 with respect to the peak and the total flux respectively. 

The flux normalized to the peak of the $^{13}$CO and the C$^{18}$O in a radial cut along the PA of the disk is also shown in Fig.~\ref{continuum_vs_gas}. For comparison, the normalized flux of the continuum mm-emission is over-plotted. The peak of both lines resides inside the mm-cavity. The uncertainties for the flux are also shown and therefore the wiggles of emission beyond 0.4''  of $^{13}$CO and C$^{18}$O are within the noise of the observations and they are not significant. We do not derive the inclination and PA from the gas emission, but from the dust continuum emission. Moreover, because of the foreground absorption in some of the channels,  we do not estimate gas masses from the current observations of CO isotopologues as done by e.g. \cite{williams2014} and \cite{miotello2016}. 

In the northern component of the SR\,24 system (SR\,24N), there is no significant detection of  $^{13}$CO or C$^{18}$O  (i.e. nothing $\gtrsim 3\,\sigma$, where $\sigma$ has a value of around $\sim 1.5\,$mJy\,beam$^{-1}$ per 1.35\,km\,s$^{-1}$ channel).  \vspace{0.2cm}

\section{Data analysis}     \label{analysis}

\subsection{Disk morphology} \label{sect:morphology}

All the following analysis for fitting the morphology of the disk from the dust continuum emission is performed in the visibility domain for the two wavelengths separately. We work with each observed $(u,v)$ point since we do not assume any a-priori knowledge of the total flux, inclination, position angle, and center of the image. Hence, these are free parameters of each of the explored models ($F_{\rm{total}}$, $i$, PA, $x_0$ and $y_0$, being $x_0$ and $y_0$ the potential offset from the center taken at $\alpha_{2000}$=16:26:58.5, $\delta_{2000}$=-24:45:37.2). As a first approximation for the structure, we assume an axisymmetric disk (Fig.~\ref{continuum_observations}), and thus we focus on fitting the real part of the visibilities. The Fourier transform of a symmetric brightness distribution can be expressed in terms of the zeroth-order Bessel function of the first kind $J_0$ of the de-projected uv-distance, such that

\begin{equation}
V_{\rm{Real}} (r_{uv})=2\pi\int^\infty_0 I(r) J_0(2\pi r_{uv}r)r dr,
\label{eq:real_part}
\end{equation}

\noindent where $r_{uv}=\sqrt{u_\phi^2\cos{i}^2+v_\phi^2}$, with $u_\phi=u\cos\phi+v\sin\phi$ and $v_\phi=-u\sin\phi+v\cos\phi$, being $i$ and $\phi$ the inclination and position angle of the disk respectively.

\begin{table*}
\caption{Results from the MCMC fitting ALMA data. Ring model (Eq.~\ref{eq:ring_model})}
\label{results1}
\centering   
\begin{tabular}{c||ccccccc}
\hline
\hline       
\textbf{Data}& $R_{\rm{peak}}$ (au) &$R_{\rm{width}}$ (au)&$F_{\rm{total}}$ (Jy)&$i$ ($^\circ$)&PA ($^\circ$)&$x_0$ (mas)&$y_0$ (mas)\\[0.1cm]
\hline
$0.45$\,mm & $20.66^{+3.25}_{-3.17}$ & $36.62^{+2.36}_{-2.49}$& $1.85^{+0.16}_{-0.16}$& fixed & fixed &-$3.17^{+0.30}_{-0.17}$&$2.90^{+0.16}_{-0.34}$\\[0.1cm]
\hline
$1.30$\,mm & $42.15^{+2.29}_{-2.67}$ & $21.41^{+2.69}_{-2.04}$& $0.21^{+0.01}_{-0.02}$& $46.17^{+2.78}_{-1.02}$& $24.73^{+3.16}_{-1.31}$& $-1.40^{+0.45}_{-0.97}$ &$2.64^{+0.33}_{-0.78}$\\[0.1cm]
\hline
\end{tabular}    
\end{table*}

\begin{table*}
\caption{Results from the MCMC fitting ALMA data. Radially asymmetric ring (Eq.~\ref{eq:asymmetric_model})}
\label{results2}
\centering   
\begin{tabular}{c||cccccccc}
\hline
\hline       
\textbf{Data}&$R_{\rm{peak}}$ (au)& $R_{\rm{width}}$ (au)& $R_{\rm{width2}}$(au)&$F_{\rm{total}}$ (Jy)&$i$ ($^\circ$)&PA ($^\circ$)&$x_0$ (mas)&$y_0$ (mas)\\[0.1cm]
\hline
$0.45$\,mm &$20.88^{+2.06}_{-3.02}$ &$28.81^{+3.56}_{-2.54}$ & $35.77^{+1.32}_{-1.57}$&$1.83^{+0.19}_{-0.16}$& fixed & fixed &-$2.57^{+0.07}_{-0.08}$&$1.95^{+0.11}_{-0.45}$\\[0.1cm]
\hline
$1.30$\,mm &$37.45^{+2.85}_{-2.91}$ & $16.46^{+3.96}_{-2.94}$ & $25.00^{+1.13}_{-1.67}$&$0.21^{+0.01}_{-0.02}$& $45.76^{+2.87}_{-0.87}$& $23.63^{+2.56}_{-0.61}$& $-0.54^{+0.08}_{-0.05}$ &$1.63^{+0.14}_{-0.34}$\\[0.1cm]
\hline
\end{tabular}    
\end{table*}

\begin{table*}
\caption{Results from the MCMC fitting ALMA data. Power law + ring-Gaussian (Eq.~\ref{eq:power_model})}
\label{results3}
\centering   
\begin{tabular}{c||cccccccc}
\hline
\hline       
\textbf{Data}&$R_{\rm{peak}}$ (au)& $R_{\rm{width}}$ (au)& $\gamma$&$F_{\rm{total}}$ (Jy)&$i$ ($^\circ$)&PA ($^\circ$)&$x_0$ (mas)&$y_0$ (mas)\\[0.1cm]
\hline
$0.45$\,mm &$21.29^{+2.27}_{-1.62}$ & $36.06^{+1.51}_{-1.25}$ &$1.64^{+0.27}_{-0.29}$&$1.87^{+0.19}_{-0.17}$& fixed & fixed &-$4.44^{+0.23}_{-0.54}$&$2.85^{+0.32}_{-0.76}$\\[0.1cm]
\hline
$1.30$\,mm &$43.98^{+1.79}_{-1.46}$ & $22.74^{+2.91}_{-1.96}$ &$1.34^{+0.24}_{-0.29}$&$0.21^{+0.02}_{-0.01}$& $46.31^{+1.88}_{-1.07}$& $24.30^{+2.14}_{-0.86}$& $-1.21^{+0.28}_{-0.45}$ &$1.47^{+0.39}_{-0.53}$\\[0.1cm]
\hline
\end{tabular}    
\tablecomments{The range given for each parameter corresponds to the 95\% credible range. The PA and $i$ are fixed for the fitting of 0.45\,mm data and taken to be the same values found from the fitting of the 1.3\,mm data (specifically PA=24$^\circ$ and $i=46^\circ$}).
\end{table*}

To fit the morphology of the disk, because the visibilities and continuum maps reveal a cavity at the two wavelengths, we explore models where the intensity profile has a ring shape. The fitting is conducted using Markov chain Monte Carlo (MCMC) method. The first model we use is a radially symmetric Gaussian ring, with two extra free parameters, for a total of seven free parameters($R_{\rm{peak}}, R_{\rm{width}}$, $F_{\rm{total}}$, $i$, PA, $x_0$ and $y_0$), such that the intensity radial profile is given by 

\begin{equation}
I(r)=C\exp\left(-\frac{(r-R_{\rm{peak}})^2}{2R_{\rm{width}}^2}\right),
\label{eq:ring_model}
\end{equation}

\noindent where the constant $C$ is related with the total flux of the disk as 

\begin{equation}
C=\frac{F_{\rm{total}}}{\int^\infty_0 I(r) J_0(0)r dr}. 
\end{equation}

The cuts along the PA of the disk in Fig.~\ref{continuum_observations} show that the ring is not necessarily a symmetric Gaussian around the peak in the radial direction (in particular for the 0.45\,mm emission), and hence we use two other different models to mimic a radially asymmetric ring (still azimuthally symmetric since our models are focused on fitting the real part of the visibilities). In the first of these models we assume an asymmetric Gaussian with two different widths that coincide at the location of the peak of emission, such that 

\begin{equation}
I(r)=\left\{ \begin{array}{rcl}
C\exp\left(-\frac{(r-R_{\rm{peak}})^2}{2R_{\rm{width}}^2}\right) &\mbox{for} & r\leq R_{\rm{peak}}\\
C\exp\left(-\frac{(r-R_{\rm{peak}})^2}{2R_{\rm{width2}}^2}\right) &\mbox{for} & r> R_{\rm{peak}},
\end{array}\right.
\label{eq:asymmetric_model}
\end{equation}

\noindent this model has a total of eight free parameters: $R_{\rm{peak}}, R_{\rm{width}},$ $R_{\rm{width2}}$, $F_{\rm{total}}$, $i$, PA, $x_0$ and $y_0$. 
This radially asymmetric ring model is also motivated by results of particle trapping in radial pressure bumps. These models of dust evolution predict that the regions where dust accumulates become narrower for larger grains (and therefore for longer wavelengths). Additionally, it is expected that the accumulation is radially narrower at longer times of evolution ($\gtrsim1\,$Myr), because micron-sized dust particles require long time to grow  to mm-sizes in the outer parts of the disk, from where they will then drift towards the pressure bump. At longer times ($\sim5$\, Myr), the emission from the models is expected to be a symmetric ring. As a consequence, the morphology of the trapped dust is expected to be an asymmetric ring in the radial direction (which can be mimicked assuming $R_{\rm{width}}<R_{\rm{width2}}$ in Eq.~\ref{eq:asymmetric_model}) at shorter times after the pressure bump is formed ($\lesssim1\,$Myr), becoming narrower and radially symmetric at longer times  \citep[$\sim5\,$Myr, see e.g. Fig.\,4 from][]{pinilla2015_alma}.

Secondly, we assume a combination of a Gaussian profile with a power-law. This also has eight free parameters, namely $R_{\rm{peak}}, R_{\rm{width}}$, $\gamma$,  $F_{\rm{total}}$, $i$, PA, $x_0$ and $y_0$, given by

\begin{equation}
I(r)=C\left[r^{-\gamma}+\exp\left(-\frac{(r-R_{\rm{peak}})^2}{2R_{\rm{width}}^2}\right) \right].
\label{eq:power_model}
\end{equation}

The motivation of this model is to investigate the potential emission from the inner disk and its possible dependency with wavelength. 

For the fits, we used {\it emcee} \citep{foreman2013}, which allows us to efficiently sample the parameter space in order to maximize the likelihood result for each model. Maximizing the likelihood function ($\mathscr{L}(\Theta)$) is equivalent to minimize the negative of the logarithm of the likelihood, since the logarithm is an increasing function over the entire range. Therefore we aim to minimize the following function
 
\begin{eqnarray}
-\log(\mathscr{L}(\Theta))&=&-\frac{1}{2}\sum_{i=1}^{n}[\log(2\pi\sigma_i^2) \nonumber\\
&+&\frac{(V_{\rm{Real, obs}}^i-V_{\rm{Real, model}}^i)^2}{2\sigma_i^2}]
\label{eq:likelihood}
\end{eqnarray}

\noindent where $\sigma$ is the uncertainty of each observed $(u,v)$ point, $n$ is the total number of data, $V_{\rm{Real, obs}}$ is the real part of the observed visibilities, and  $V_{\rm{Real, model}}$ are the visibilities for each model calculated with Eq.~\ref{eq:real_part}. We adopted a set of uniform prior probability distributions for  the free parameters explored by the Markov chain in the three models, specifically: 

\begin{eqnarray}
R_{\rm{peak}} &\in& [1, 100]\,\rm{au} \nonumber\\
R_{\rm{width}} &\in& [1, 50]\,\rm{au}\nonumber\\
R_{\rm{width2}}&\in& [1, 50]\,\rm{au} \nonumber\\
F_{\rm{total}}&\in& [1.0, 5.0]\,\rm{Jy} \qquad \textrm{for the 0.45\,mm data}\nonumber\\ 
F_{\rm{total}}&\in& [0.02, 2.5]\,\rm{Jy} \qquad \textrm{for the 1.3\,mm data}\nonumber\\ 
i&\in& [10, 80]\,^\circ\nonumber\\
\rm{PA}&\in& [10, 80]\,^\circ\nonumber\\
x_0&\in& [-0.2, 0.2]\,''\nonumber\\
y_0&\in& [-0.2, 0.2]\,''\nonumber\\
\gamma&\in& [-3, 3].
\label{eq:parameter_space}
\end{eqnarray}

For the radial grid, we assume $r\in[1-500]\,$au with steps of 0.1\,au.  The burn-in phase for convergency is $\sim1000$ steps, which is $\sim10$ times the autocorrelation time of 100 steps \citep[e.g.][]{sokal1994, tazzari2016}. We let the Markov chain to sample the parameter space for another thousands of steps, for a total of 4000 steps with 1000 walkers. Each measurement set is fitted separately and therefore we used  {\it emcee} to fit a total of six models (3 models for 2 different data sets). To simplify the fitting process, we obtained the PA and the $i$ using the 1.3\,mm data (since it has better signal-to-noise)  for each model, and keep the best-fit values of these two parameters to fit the 0.45\,mm data.

\begin{figure*}
 \centering
 \tabcolsep=0.1cm 
   \begin{tabular}{cc}   
   	\includegraphics[width=9cm]{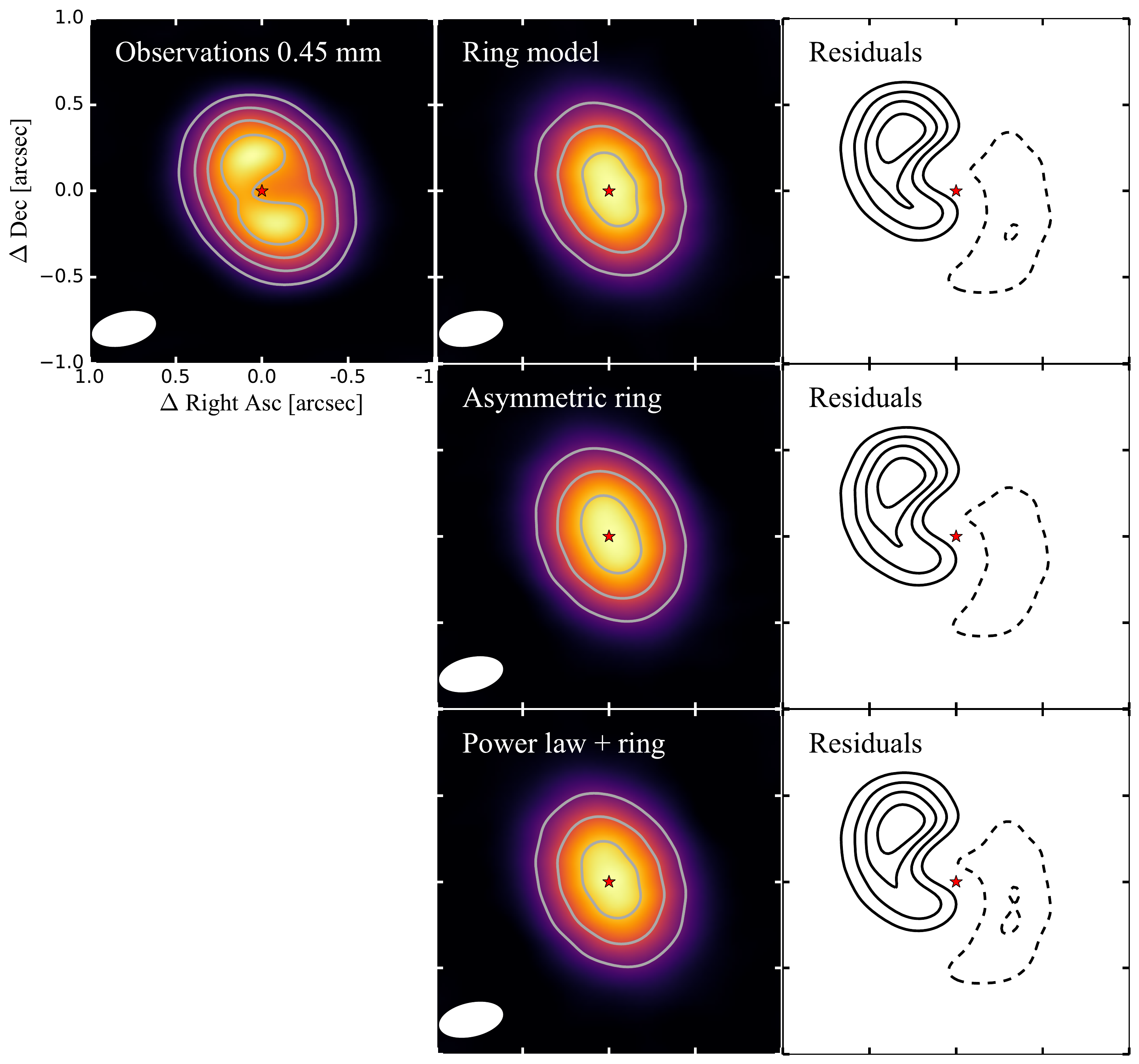}&\includegraphics[width=9cm]{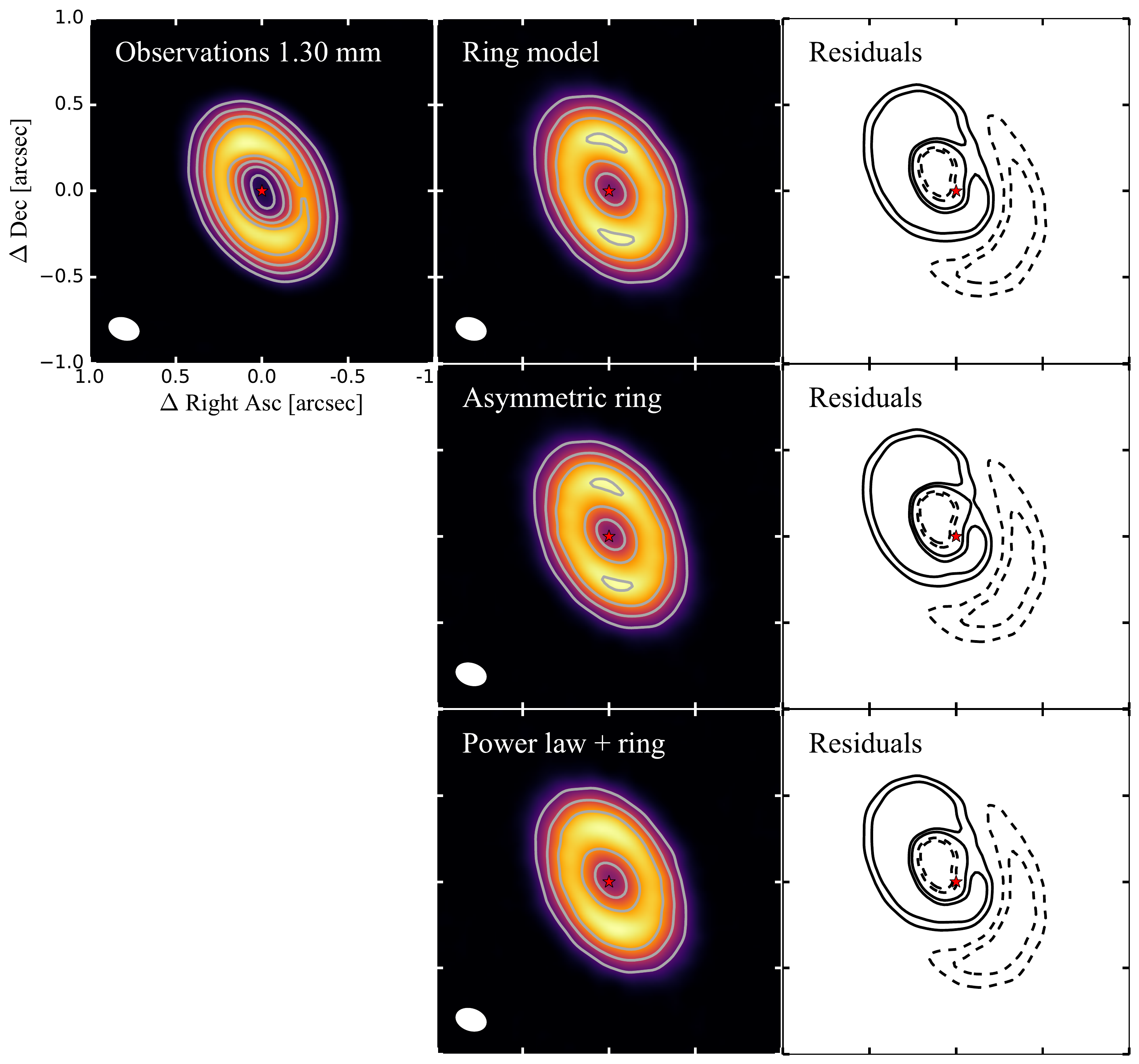}
   \end{tabular}
    \caption{Observations, best-fit models, and residuals for the 0.45\,mm data \emph{(left panels)} and the 1.3\,mm data\emph{(right panels)}. In the observations and model maps the contours are every 20, ..., 100\% of the peak value (Table~\ref{tbl:obsproperties}). In the residual maps (data-model),  the contours are every $\pm3\sigma$ (from -12$\sigma$ to 12$\sigma$ for the 0.45\,mm model and from -6$\sigma$ to 6$\sigma$ for the 1.3\,mm model), showing the negative contours in dashed lines. }
   \label{model_residuals}
\end{figure*}

\begin{figure*}
 \centering
 \tabcolsep=0.05cm 
   \begin{tabular}{ccc}
   	\includegraphics[width=6cm]{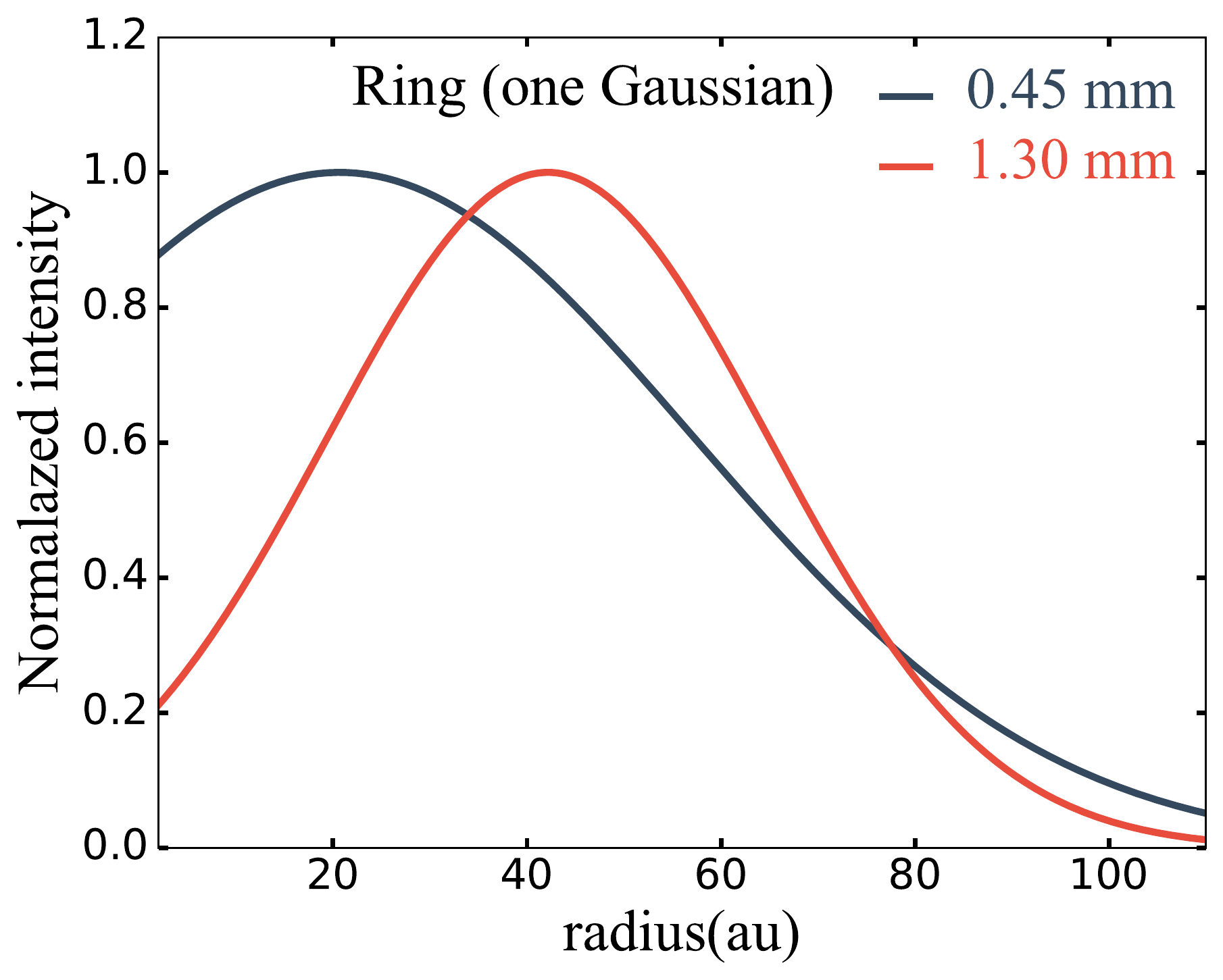}&
	\includegraphics[width=6cm]{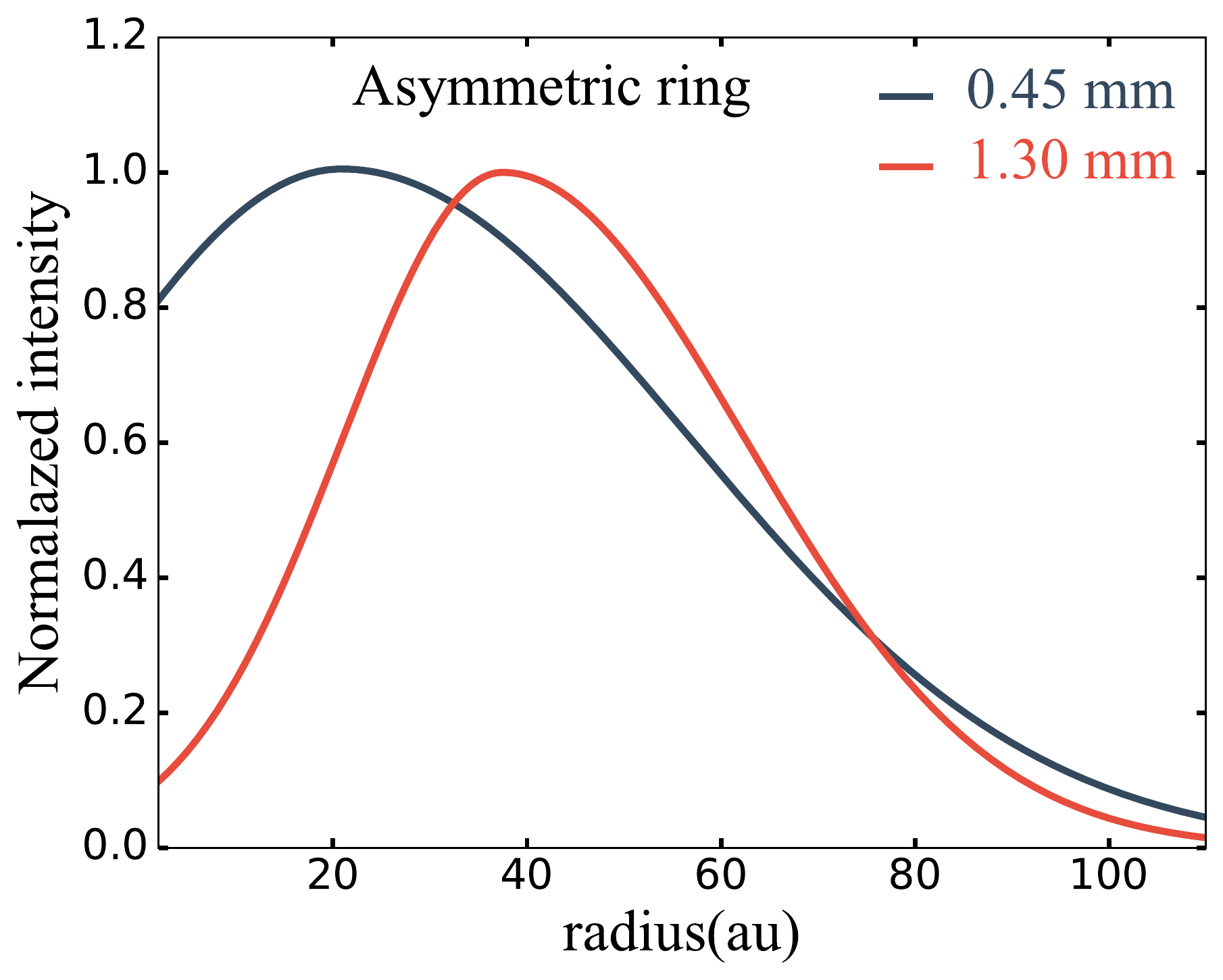}&
   	\includegraphics[width=6cm]{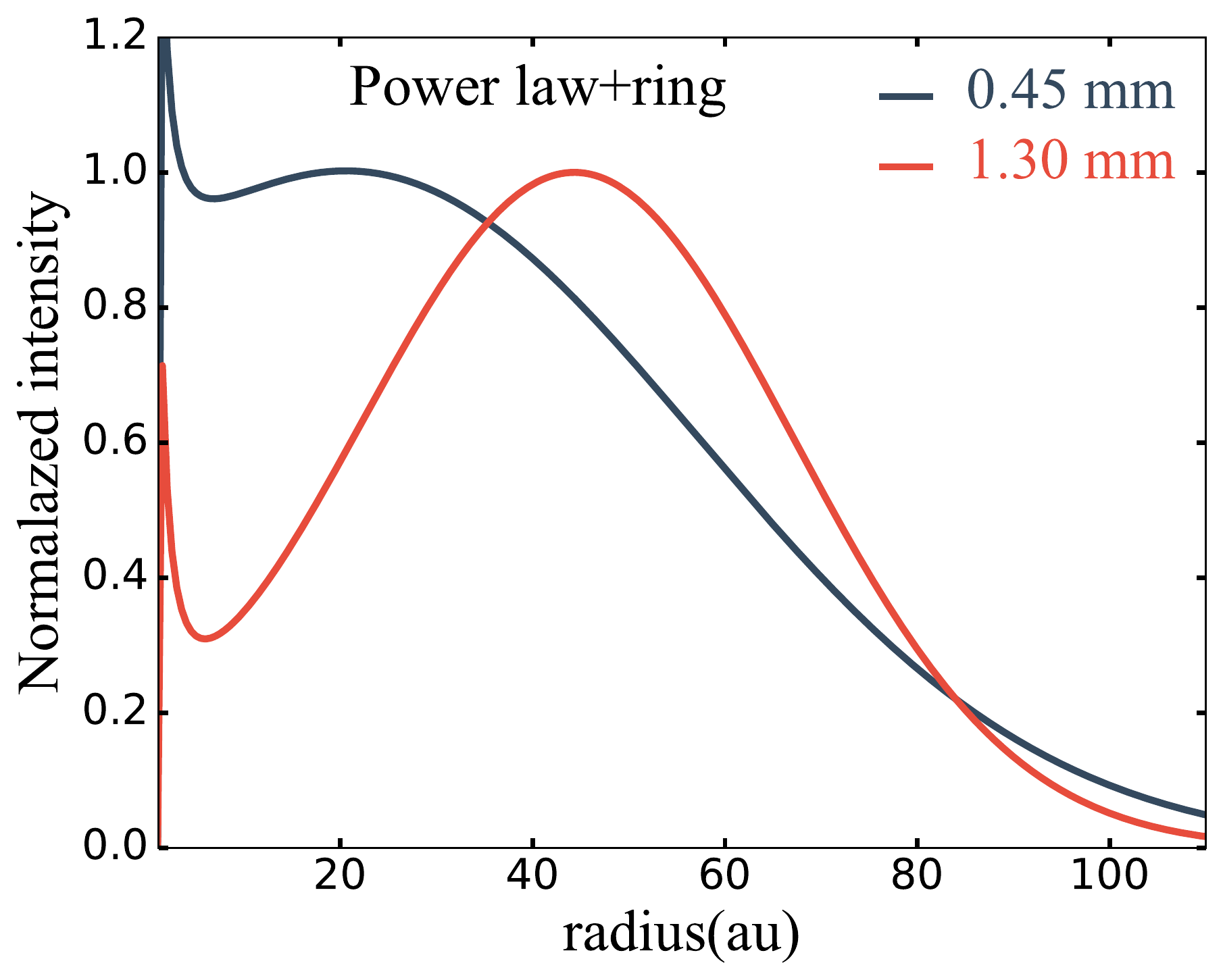}   
	\end{tabular}
   \caption{Best fit models (from left to right: ring, asymmetric ring and power law+ ring) for the 0.45 and 1.3\,mm data.  For each case, the intensity is normalized to the value at the location of peak of the ring.}
   \label{best_models_intensity}
\end{figure*}

The results are summarized in Tables~\ref{results1},~\ref{results2},~\ref{results3}, and Fig.~\ref{best_models_intensity}. All three models show that the peak of the ring is located further out and becomes narrower at longer wavelength, with a  difference of around $\sim$20\,au for the location of the peak and $\sim$10\,au for the width(s). The models of the radially asymmetric ring, shows that  $R_{\rm{width}}<R_{\rm{width2}}$, creating a slightly (radially) asymmetric ring with an outer tail. The model of the power law together with a Gaussian gives a slightly stepper profile for the 0.45\,mm emission. The inclination and position angle obtained with the three models give very similar values, with mean values of 46$^\circ$ and 24$^\circ$ respectively, in agreement with the values found by \cite{andrews2011} and \cite{marel2015} and in Sect.~\ref{observations}. The values obtained for the shift of the center are very low compared with pixel size from the observations (0.02'' for the 1.3\,mm observations and 0.04'' for the 0.45\,mm observations). 

We image the models and residuals  (data-model) using identical $(u,v)$ coordinates as the actual ALMA observations. Figure~\ref{model_residuals} shows the synthetic images of the best fit models and the corresponding residuals. In general, the quality of the three fits is similar.  For the 1.3\,mm data, all three models reproduce roughly the same amount of residuals with respect to the rms of the observations. This is because all three models resample an almost symmetric ring and a quite empty cavity, where the intensity decreases around 80-90\% with respect to the peak of emission, as the observations (left panel Fig.~\ref{continuum_observations}).  For the 0.45\,mm data, the asymmetric ring is the model that reproduces less residuals, where the emission inside the cavity only decreases by $\sim20$\% with respect to the peak of emission.

Figure~\ref{best_models_intensity} shows the profile of the normalized intensity with respect to peak value, calculated with the best fit parameters (Tables~\ref{results1},~\ref{results2}, and ~\ref{results3}), for each case and each wavelength (for this plot, the inclination and the position angle are taken to have the same value for all three models, that is the mean value, $i=46^\circ$ and PA$=24^\circ$).  All of these thee models resemble a roughly symmetric ring-like emission at the two wavelengths, and the profiles in Fig.~\ref{best_models_intensity} are comparable with the azimuthally  averaged radial profile of the de-projected images (left panel in Fig.~\ref{spectral_index}).

Independently of the model, an asymmetric structure persists in the maps of the residuals  (Fig.~\ref{model_residuals}). The model under-predicts the flux in the north-east and over-predicts the flux in the south-west with similar magnitude and morphology. The residuals show a spiral-shape like structure similar to those found in the residual maps of the visibilities analysis of SR\,21 and HD\,135344B \citep[][]{perez_L2014, pinilla2015_alma}. The residuals peak along $\sim47\pm2$. Taking a radial cut along this angle, the positive residuals peak around $0.45\pm{0.1}''$  from the center, and the dip of emission has its minimum around the same location in the opposite side of the disk. The residual map of the 1.3\,mm data shows an additional structure at $\sim0.15''$ with a swap in the flux emission (negative in the north-east and positive in the south-west). As an experiment, we also performed the MCMC fitting keeping PA, $i$, $x_0$, and $y_0$ fixed and assuming the values obtained with {\tt uvmodelfit}. In this experiment, we found a higher amount of residuals, but with similar shape as the ones shown in Fig.~\ref{model_residuals}. In our analysis, we only fitted the real part of the visibilities, assuming none azimuthal variations.  As a result, if an asymmetry arises due to an offset, the model optimizes the fit towards a symmetric emission. Hence, in the framework of our models, it is difficult to conclude how significant is the amount of residuals and the their shape and higher angular resolution observations are required to confirm if these potential asymmetries are real.

\cite{zhang2016} modeled the visibility profile of several disks with multiple rings (such as HL\,Tau, TW\,Hya and HD\,163296). Applying their method to our data does not seem suitable since there is not more than one distinctive peak in the visibility profile. However, as a test we did an experiment of fitting the visibilities at 1.3\,mm with at least two rings. In this case, the fit converges to one single ring that dominates the intensity profile. More complex models that include asymmetric structures, such as spiral arms, may improve the fit (Fig.~\ref{model_residuals}). However, due to the high degeneracy of including several parameters to fit both the imaginary and the real part of the visibilities simultaneously, we do not perform such analysis. 

\subsection{Optical thickness and spectral index interpretation} \label{sect:spectral_index}

The slope of the spectral energy distribution (SED) at millimeter wavelengths, or spectral index ($\alpha_{\rm{mm}}$, such that $F_{\rm{mm}}\propto\nu^{\alpha_{\rm{mm}}}$), has been widely used to trace millimeter-grains in protoplanetary disk. If the millimeter emission is optically thin, low values of the spectral index ($\lesssim3.5$) indicate the growth of particles to millimeter sizes. The spatially integrated spectral index ($\alpha_{\rm{mm}}$) in protoplanetary disks observed in different star forming regions and around different stellar types has values lower than 3.5 \citep[e.g.][]{birnstiel2010, testi2014}. Spatial variations of the spectral index in different protoplanetary disks have been resolved, where in most of the cases the spectral index increases radially, evidencing that the grain size decreases for increasing radius  \citep[e.g.][]{banzatti2011, perez_L2012, perez_L2015, trotta2013,  tazzari2016}, as expected from radial drift, as seen in dust evolution models \citep[e.g.][]{birnstiel2012}.

However, for transition disks the spectral index is expected to increase towards the outer edge of the cavity, that is, towards the location where particles are trapped and have grown to mm-sizes. For these disks, the spatially integrated spectral index is also expected to be higher for larger cavities \citep{pinilla2014}. There are  few transition disks where the spectral index has been imaged, HD\,142527, IRS\,48 and SR\,21 \citep{casassus2015, marel2015_irs48, pinilla2015_alma}, and in these few cases the spectral index decreases towards the location where the mm-emission peaks and where a particle accumulation is expected. 

\begin{figure*}
 \centering
  \tabcolsep=0.1cm 
   \begin{tabular}{ccc}   
   \includegraphics[width=6cm]{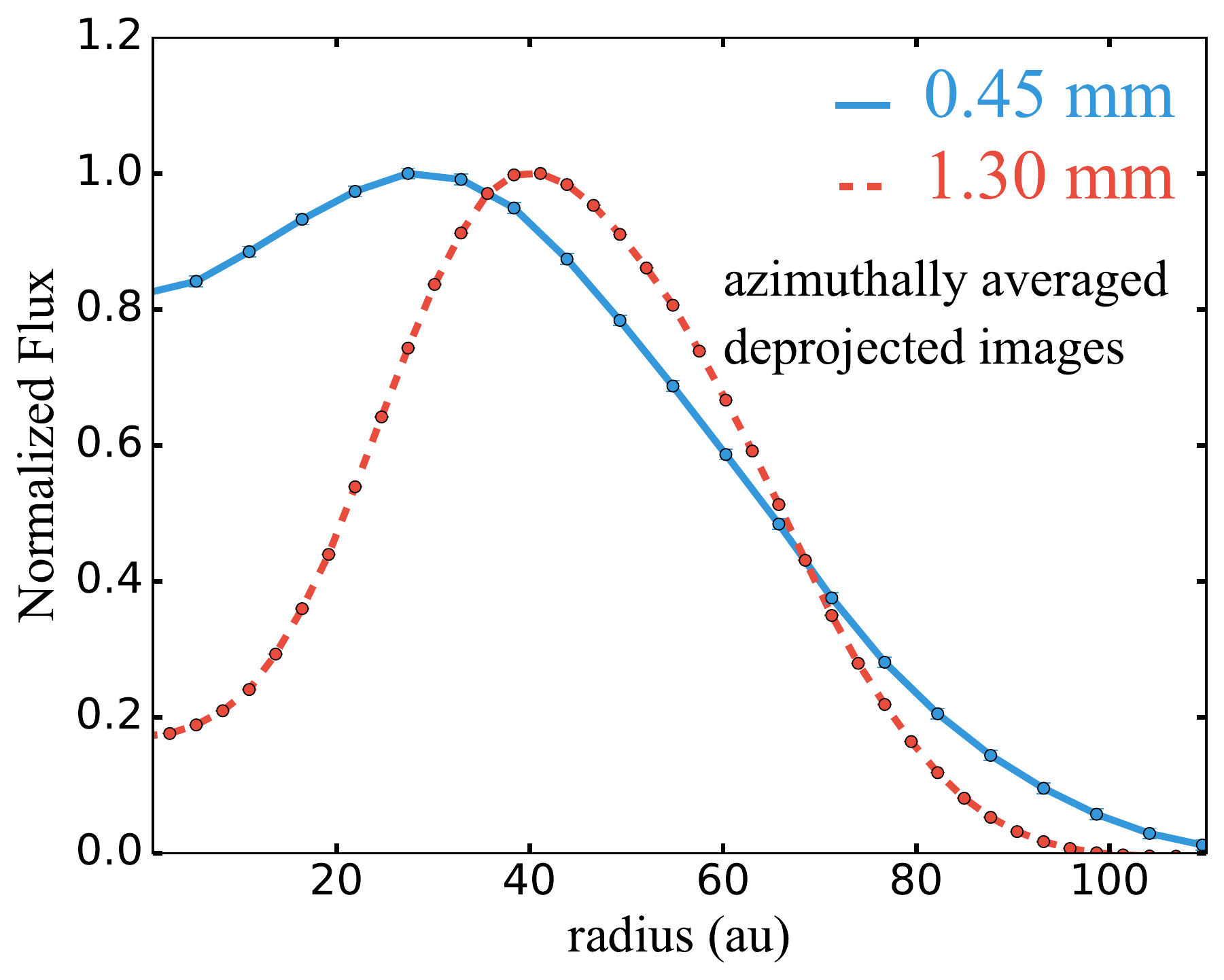}&
   \includegraphics[width=6cm]{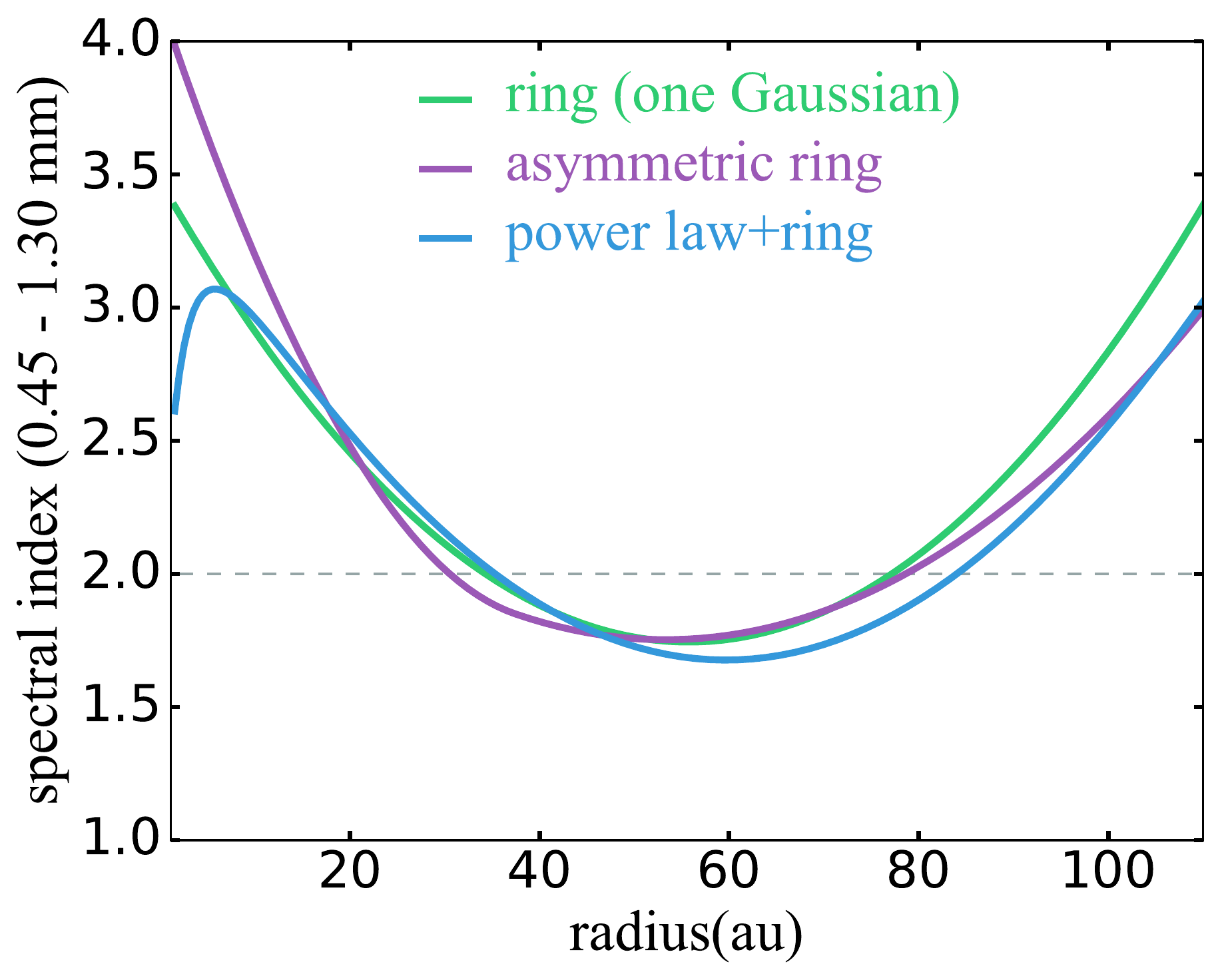}&
   \includegraphics[width=6cm]{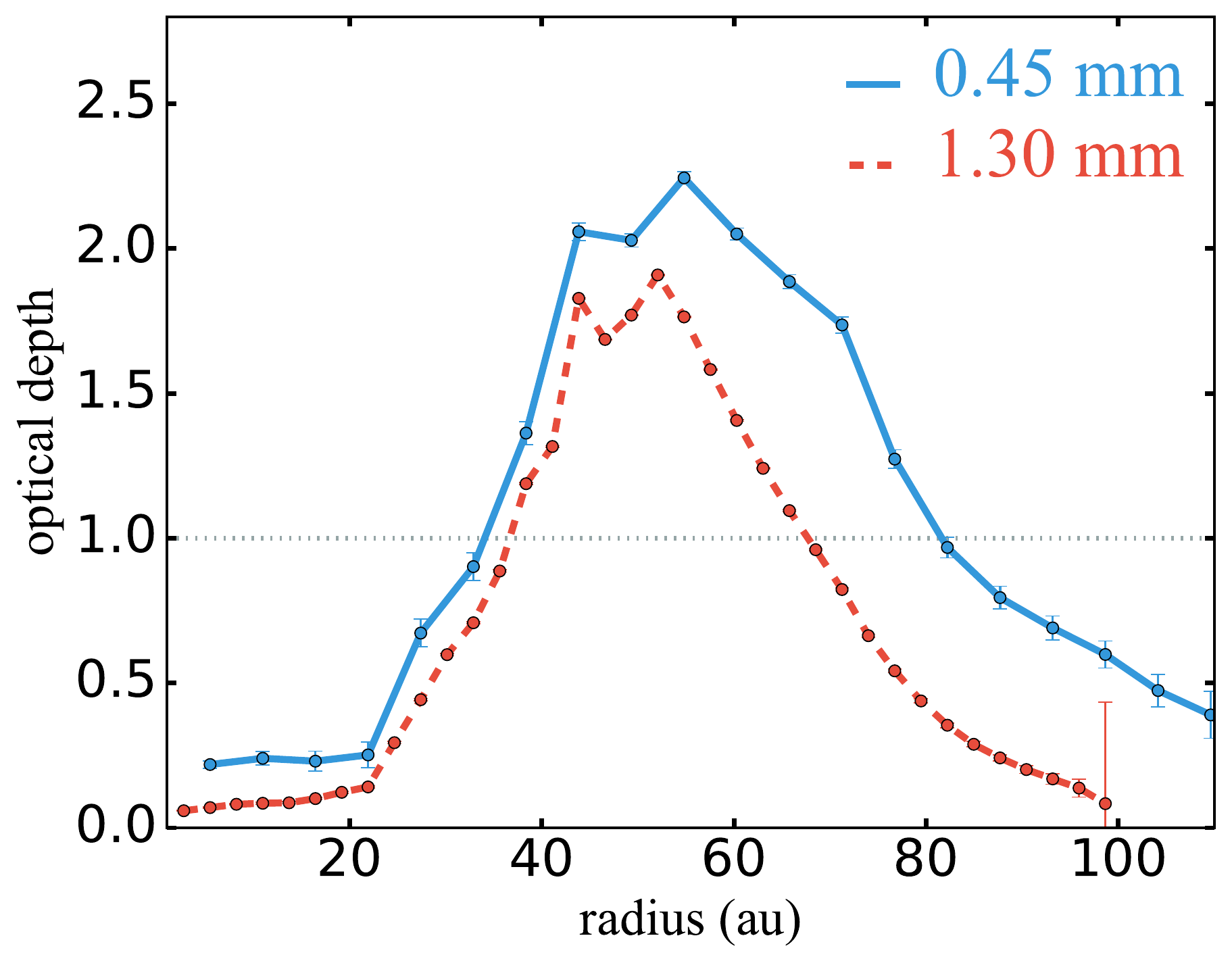}
   \end{tabular}
    \caption{\emph{Left panel: } Normalized flux at 0.45 and 1.3\,mm of the azimuthally averaged de-projected ALMA images. The error bars are of the size of the points \emph{Middle panel:} Spectral index calculated from the best-fit parameters of the intensity profiles at each wavelengths (Sect.~\ref{sect:morphology}) \emph{Right panel:} Optical depth obtained by calculating the brightness temperature from the continuum images and by assuming the physical temperature from \cite{marel2015}.  The error bars are included and obtained from error propagation taking into account the rms and the standard deviation from the azimuthally averaged flux. }
   \label{spectral_index}
\end{figure*}

The middle panel of Fig.~\ref{spectral_index} shows the radial profile of the  spectral index calculated from the intensity profiles taking the best fit parameters for each model at each wavelength described in Sect~\ref{sect:morphology}.  At the location of the ring, the spectral index has values lower than 2.0, specifically from $32\pm3$\,au to $81\pm4$\,au. The right panel of Fig.~\ref{spectral_index} also shows the optical depth $\tau$ obtained from the brightness temperature, which is calculated from the azimuthally averaged flux of the de-projected image (displayed in the left panel of Fig.~\ref{spectral_index}) and  without assuming the Rayleigh-Jeans regime. For the physical temperature, we assume the mid-plane values  from dust radiative transfer models for SR\,24S in \cite{marel2015}. The error bars are obtained from error propagation taking into account the rms of the observations and the standard deviation from the azimuthally averaged flux values. With the assumed temperature, the emission is optically thick at both wavelengths ($\tau>1$) close to the location of the ring-like emission. In particular, $\tau>1$ from $\sim$33 to $\sim$81\,au for the 0.45\,mm emission and from  $\sim$36 to $\sim$66\,au for the 1.3\,mm emission. Therefore, the decrease of the spectral index towards the location of the ring is likely because of optical thickness and with the current observations,  $\alpha_{\rm{mm}}$ cannot be interpreted as  grain growth inside the ring.

Observations at shorter (optically thick) wavelengths can provide information of the temperature distribution in SR\,24S; and at longer (optically thin) wavelengths, which can give direct information of the dust size, are therefore needed to better constrain the dust density distribution in SR\,24S.  

\section{Discussion}     \label{discussion}

Embedded planets inside the mm-cavities are commonly used to explain the observed gas and dust structures of transition disks. From planet-disk interaction together with dust evolution models, it is expected that at the outer edge of the planet-carved gap, the dust particles accumulate and grow to mm-sizes. For the gas, it is expected that the cavity is smaller and less depleted than the mm-cavity, as observed in several transition disks \citep[e.g.][]{marel2016}. The small grains (micron-sized) particles are also expected  to have a different distribution than the mm-dust, with smaller or no-gaps, depending on planet mass and disk viscosity \citep[e.g.][]{maria2016}. 

From the current observations of  $^{13}$CO and C$^{18}$O in the transition disk SR\,24S, the gas emission is centrally peaking inside the mm-dust cavity. It is possible that the gas cavity (or gap) remain unresolved or that the gas is not depleted inside the mm-cavity. The latter case suggests that if embedded planets are the cause of the mm-cavity, they must be low-mass planets \citep[$\lesssim 0.1-1\,M_{\rm{Jup}}$, see e.g.][]{pinilla2012, zhu2012, rosotti2016}. An alternative explanation is dust trapping at the outer edge of a dead zone. In this scenario, strong accumulation of millimeter and centimeter-sized particles are expected at the location of the outer edge of the dead zone, while the gas is only slightly depleted in the inner part of the disk \citep{pinilla2016b}.  Information of the scattered light emission of SR\,24S can give important insights to distinguish between the two possibilities, since in the dead zone scenario the cavity size in scattered light is expected to have the same size as the mm-cavity, contrary to the planet-disk interaction scenario  \citep{pinilla2016b}. The resolution of previous near-infrared observations was not enough to detect a potential cavity in small grains in SR\,24S \citep{mayama2010}. Additional information about the distribution of the intermediate grain sizes, which can be obtained by  millimeter-wave polarization can also give significant insights to recognize whether or not planet-disk interaction is the most likely cause for the origin of the cavity in SR\,24S \citep{pohl2016}. Models of internal photoevaporation predict a cavity in gas \emph{and} dust  (an low accretion rates) and therefore this scenario can be ruled out as a potential origin for the mm-cavity in SR\,24S \citep{owen2012}.

Our analysis for the morphology of the dust emission at 0.45 and 1.30\,mm in SR\,24S shows that at longer wavelengths the ring-like emission becomes narrower. This is in agreement with models of dust trapping in a pressure bump, since larger grains that are traced with longer wavelengths accumulate more efficiently at the location of the pressure maximum (or particle trap). Contrary to model predictions, in the observations there is also a shift of the peak of emission. This shift may result from optically thick emission at the two wavelengths and observations at longer (optically thin) wavelengths are required to see if the peak of emission for larger grains shifts. 

The models of particle trapping by planets predict that at early times of evolution ($\sim1\,$Myr), the ring like emission is a radially asymmetric ring,  where the outer width of the ring is expected to be higher than the inner width of the ring \citep[e.g.][]{pinilla2015_alma}. As a result of slow grain growth in the outer disk, the intensity profile at early times of evolution is a ring with an outer tail. At longer times ($\sim5$\, Myr), the emission from the models is expected to be a symmetric ring. The dust morphology found from the current observations of SR\,24S shows that the emission at both wavelengths is well represented by an almost symmetric ring. The models with the radially asymmetric Gaussian give similar profiles than the perfect symmetric ring (Fig.~\ref{best_models_intensity}). This suggests that if trapping is occurring due to embedded planets, the trapping process has occurred already for long times of evolution, which is in agreement with the age reported by \cite{wilking2005} of $\sim4$\,Myr for SR\,24S.

In addition, the emission inside the mm-cavity at 0.45\,mm is less depleted than at 1.3\,mm. This could happen because smaller grains may not be completely filtered at the outer edge of the planet induced-gap, which allows that small grains pass through the gap and replenish the inner disk. The SED of SR\,24S shows near-infrared (NIR) excess emission \citep[e.g.][]{andrews2011}, which suggests the presence of an inner disk. The combination of NIR excess and ring-like emission can be reproduced in models of partial filtration at the outer edge of gap opened by a massive planet, with changes of the dust dynamics near the water snow line. Without variations of the fragmentation velocities of the particles near the snow line, no NIR excess is produced because even in the case of partial filtration, when small grains pass the planet gap, the growth in the inner disk is so efficient that the grains are lost in very short timescales ($\lesssim0.1\,$Myr). The required conditions to keep a long-lived inner disk and an outer ring are in agreement with the gas emission peaking inside the mm-cavity that suggests the presence of low mass planet(s), allowing the inner disk to be continuously replenished by the outer edge, even at very long times of evolution \citep{pinilla2016a}.

The analysis of the current data suggest that the morphology of SR\,24S disk is not completely azimuthally symmetric and complex asymmetric structures may exist. These asymmetric structures can also be linked with the presence of planets, MRI effects, or from the interaction with the binary system SR\,24N.  Higher angular resolution observations are needed to confirm the existence of such structures and to better understand their nature and origin. 

The current observations of SR\,24N suggest that this disks is poor on mm-grains and cold molecular gas. A possible explanation is that the growth of sub-micron sized particles to larger objects is inhibited by the high dust relative velocities of collisions expected around binary systems \citep{zsom2011}. CIAO observations at H-band showed that the disks in scattered light seem to be extended enough to fill the effective Roche radius of the system, probing tidal interactions between the two disks \citep{mayama2010}. For the gas, it is possible that only warm gas inside the truncation radii is present \citep{brown2013}. The truncation radii for close binaries is expected to be $\lesssim0.4a$ \citep[specific value depends on the eccentricity and inclination, e.g.,][]{artymowicz1994, miranda2015}, being $a$ the separation of the two stars. In this case, $a\sim0.2''$, which means that the truncation radii is around 0.08'' ($\sim10$\,au), which is smaller than the current resolution of the ALMA observations. An alternative possibility for the lack of mm-dust and gas in SR\,24N is that planets already formed in this binary system, depleting the  disk in gas and large grains.

\section{Conclusions}     \label{conclusion}

We present ALMA observations of the transition disk SR\,24S at 0.45 and 1.3\,mm respectively. Our findings are as follows:

\begin{enumerate}
\item The visibilities and images of the continuum emission at the two wavelengths are well described by ring-like emission. The width of such ring is narrower at longer wavelengths in agreement with particle trapping in pressure bumps. The ring is mostly radially symmetric suggesting that if the trapping process is due to embedded planets, it must occur for long times of evolution.

\item  Inside the mm-cavity, the emission at 0.45\,mm is less reduced than at 1.3\,mm ($\sim$20\% vs. 85\% of depletion with respect to the peak of emission). This could be linked with a constant replenishment of small grains from the outer disk (or partial filtration of particles at the outer edge of a gap). This is agreement with the NIR excess of the SED of SR\,24S, i.e. with the presence of an inner disk in the first au. 

\item Analysis in the visibility domain reveal a complex morphology of the disk with potential asymmetric shape. Higher angular resolution observations are required to confirm such structure(s).

\item We detect $^{13}$CO and C$^{18}$O (J=2-1) emission in SR\,24S disk. The current observations show that the emission of both molecular lines peak at the center of the mm-cavity, in contrast with most of the transition disks observed so far with ALMA. This suggest that  whatever is the origin of mm-cavity allows enough gas to reside in the inner part of the disk, as in the case of dead zones \citep{pinilla2016b}. Internal photoevaporation is unlikely as a potential origin for the seen structures in SR\,24S given the high accretion rate, the large mm-cavity and the gas emission peaking in the inner disk.

\item There is no detection of the dust continuum, $^{13}$CO, or C$^{18}$O emission at the location of the northern component (SR\,24N) of this hierarchical triple system SR\,24.  This suggests that in SR\,24N, either planets already formed in this binary system, depleting this circumbinary disk in gas and mm-grains; or dust growth to mm-sizes is inhibited in this disk and that only warm gas inside the truncation radii of the binary is present. 

\item Assuming physical temperatures from previous modeling for SR\,24S disk, we conclude that the emission at both wavelengths is optically thick close to the location of the ring-like emission. High angular resolution observations at longer wavelength are reeded to investigate potential spatial changes (radial and/or azimuthal) of the dust density distribution through spatially resolved spectral index variations. \vspace{0.3cm}

\end{enumerate}

\acknowledgements{
The authors are thankful with the ALMA contact and data reducer, Luke Maud, for his help to understand the data reduction and calibration. We also acknowledge Catherine Walsh, Marco Tazzari, and Sierk van Terwisga for useful discussions. P.P. acknowledges support by NASA through Hubble Fellowship grant HST-HF2-51380.001-A awarded by the Space Telescope Science Institute, which is operated by the Association of Universities for Research in Astronomy, Inc., for NASA, under contract NAS 5- 26555. Astrochemistry in Leiden is supported by the European Union A-ERC grant 291141 CHEMPLAN, by the Netherlands Research School for Astronomy (NOVA), by a Royal Netherlands Academy of Arts and Sciences (KNAW) professor prize. T.B. acknowledges funding from the European Research Council (ERC) under the European UnionÕs Horizon 2020 research and innovation programme under grant agreement No 714769. A.N. acknowledges funding from Science Foundation Ireland (Grant 13/ERC/I2907). This paper makes use of the following ALMA data: ADS/JAO.ALMA\#2013.1.00091.S and \#2011.0.00724.S. ALMA is a partnership of ESO (representing its member states), NSF (USA) and NINS (Japan), together with NRC (Canada) and NSC and ASIAA (Taiwan) and KASI (Republic of Korea), in cooperation with the Republic of Chile. The Joint ALMA Observatory is operated by ESO, AUI/NRAO and NAOJ. }


\begin{thebibliography}{}

\bibitem[Alexander et al.(2014)]{alexander2014} Alexander, R.,
  Pascucci, I., Andrews, S., Armitage, P., \& Cieza, L.\ 2014,
  Protostars and Planets VI, 475, eds. H. Beuther, R. Klessen,
  C. Dullemond, \& Th. Henning (Univ. of Arizona Press, Tucson).

\bibitem[ALMA Partnership et al.(2015)]{alma2015} ALMA Partnership, Brogan, C.~L., P{\'e}rez, L.~M., et al.\ 2015, \apjl, 808, L3 

\bibitem[Andrews \& Williams(2005)]{andrews2005} Andrews, S.~M., \&
  Williams, J.~P.\ 2005, \apjl, 619, L175 

\bibitem[Andrews et al.(2010)]{andrews2010} Andrews, S.~M., Wilner, D.~J., Hughes, A.~M., Qi, C., \& Dullemond, C.~P.\ 2010, \apj, 723, 1241 

\bibitem[Andrews et al.(2011)]{andrews2011} Andrews, S.~M., Wilner, D.~J., Espaillat, C., et al.\ 2011, \apj, 732, 42 

\bibitem[Andrews et al.(2016)]{andrews2016} Andrews, S.~M., Wilner, D.~J., Zhu, Z., et al.\ 2016, \apjl, 820, L40 

\bibitem[Artymowicz \& Lubow(1994)]{artymowicz1994} Artymowicz, P., \& Lubow, S.~H.\ 1994, \apj, 421, 651 

\bibitem[Banzatti et al.(2011)]{banzatti2011} Banzatti, A., Testi, L., Isella, A., et al.\ 2011, \aap, 525, A12 

\bibitem[Birnstiel et al.(2010)]{birnstiel2010} Birnstiel, T., Ricci, L., Trotta, F., et al.\ 2010, \aap, 516, L14 

\bibitem[Birnstiel et al.(2012)]{birnstiel2012} Birnstiel, T., Klahr, H., \& Ercolano, B.\ 2012, \aap, 539, A148 

\bibitem[Bontemps et al.(2001)]{bontemps2001} Bontemps, S., Andr{\'e},
  P., Kaas, A.~A., et al.\ 2001, \aap, 372, 173

\bibitem[Brown et al.(2013)]{brown2013} Brown, J.~M., Pontoppidan, K.~M., van Dishoeck, E.~F., et al.\ 2013, \apj, 770, 94 
 
\bibitem[Bruderer et al.(2014)]{bruderer2014} Bruderer, S., van der Marel, N., van Dishoeck, E.~F., \& van Kempen, T.~A.\ 2014, \aap, 562, A26 

\bibitem[Casassus et al.(2015)]{casassus2015} Casassus, S., Wright, C.~M., Marino, S., et al.\ 2015, \apj, 812, 126 

\bibitem[Canovas et al.(2016)]{canovas2016} Canovas, H., Caceres, C., Schreiber, M.~R., et al.\ 2016, \mnras, 458, L29 

\bibitem[Correia et al.(2006)]{correia2006} Correia, S., Zinnecker,
  H., Ratzka, T., \& Sterzik, M.~F.\ 2006, \aap, 459, 909

\bibitem[Cutri et al.(2003)]{cutri2003} Cutri, R.~M., Skrutskie, M.~F., van Dyk, S., et al.\ 2003, VizieR Online Data Catalog, 2246,  

\bibitem[de Boer et al.(2016)]{deboer2016} de Boer, J., Salter, G., Benisty, M., et al.\ 2016, \aap, 595, A114 

\bibitem[de Juan Ovelar et al.(2016)]{maria2016} de Juan Ovelar, M., Pinilla, P., Min, M., Dominik, C., \& Birnstiel, T.\ 2016, \mnras, 459, L85 

\bibitem[Dipierro et al.(2016)]{dipierro2016} Dipierro, G., Laibe, G., Price, D.~J., \& Lodato, G.\ 2016, \mnras, 459, L1 

\bibitem[Fedele et al.(2017)]{fedele2017} Fedele, D., Carney, M., Hogerheijde, M.~R., et al.\ 2017, arXiv:1702.02844 

\bibitem[Flock et al.(2015)]{flock2015} Flock, M., Ruge, J.~P., Dzyurkevich, N., et al.\
2015, \aap, 574, A68 

\bibitem[Foreman-Mackey et al.(2013)]{foreman2013} Foreman-Mackey, D., Hogg, D.~W., Lang, D., \& Goodman, J.\ 2013, \pasp, 125, 306 

\bibitem[Ghez et al.(1993)]{ghez1993} Ghez, A.~M., Neugebauer, G., \&
  Matthews, K.\ 1993, \aj, 106, 2005

\bibitem[Ginski et al.(2016)]{ginski2016} Ginski, C., Stolker, T., Pinilla, P., et al.\ 2016, \aap, 595, A112 

\bibitem[Greene et al.(1994)]{greene1994} Greene, T.~P., Wilking,
  B.~A., Andre, P., Young, E.~T., \& Lada, C.~J.\ 1994, \apj, 434, 614 

\bibitem[Hildebrand(1983)]{hildebrand1983} Hildebrand, R.~H.\ 1983, \qjras, 24, 267 
 
\bibitem[Loinard et al.(2008)]{loinard2008} Loinard, L., Torres, R.~M., Mioduszewski, A.~J., \& Rodr{\'{\i}}guez, L.~F.\ 2008, \apjl, 675, L29 

\bibitem[Mamajek(2008)]{mamajek2008} Mamajek, E.~E.\ 2008, Astronomische Nachrichten, 329, 10 

\bibitem[Mayama et al.(2010)]{mayama2010} Mayama, S., Tamura, M., Hanawa, T., et al.\ 2010, Science, 327, 306

\bibitem[Miotello et al.(2016)]{miotello2016} Miotello, A., van Dishoeck, E.~F., Kama, M., \& Bruderer, S.\ 2016, \aap, 594, A85 

\bibitem[Miranda \& Lai(2015)]{miranda2015} Miranda, R., \& Lai, D.\ 2015, \mnras, 452, 2396 

\bibitem[Natta et al.(2006)]{natta2006} Natta, A., Testi, L., \& Randich, S.\ 2006, \aap, 452, 245 

\bibitem[Nuernberger et al.(1998)]{nuernberger1998} Nuernberger, D., Brandner, W., Yorke, H.~W., \& Zinnecker, H.\ 1998, \aap, 330, 549 

\bibitem[Owen \& Clarke(2012)]{owen2012} Owen, J.~E., \& Clarke, C.~J.\ 2012, \mnras, 426, L96 

\bibitem[P{\'e}rez et al.(2012)]{perez_L2012} P{\'e}rez, L.~M., Carpenter, J.~M., Chandler, C.~J., et al.\ 2012, \apjl, 760, L17 

\bibitem[P{\'e}rez et al.(2014)]{perez_L2014} P{\'e}rez, L.~M., Isella, A., Carpenter, J.~M., \& Chandler, C.~J.\ 2014, \apjl, 783, L13 

\bibitem[P{\'e}rez et al.(2015)]{perez_L2015} P{\'e}rez, L.~M., Chandler, C.~J., Isella, A., et al.\ 2015, \apj, 813, 41 

\bibitem[P{\'e}rez et al.(2016)]{perez_L2016} P{\'e}rez, L.~M., Carpenter, J.~M., Andrews, S.~M., et al.\ 2016, Science, 353, 1519 

\bibitem[Perez et al.(2015)]{perez_s2015} Perez, S., Casassus, S., 
M{\'e}nard, F., et al.\ 2015, \apj, 798, 85

\bibitem[Pinilla et 
al.(2012b)]{pinilla2012} Pinilla, P., Benisty, M., \& Birnstiel, T.\
2012b,A\&A, 545, A81

\bibitem[Pinilla et al.(2014)]{pinilla2014} Pinilla, P., Benisty, M., Birnstiel, T., et al.\ 2014, \aap, 564, A51 

\bibitem[Pinilla et al.(2015)]{pinilla2015_alma} Pinilla, P., van der Marel, N., P{\'e}rez, L.~M., et al.\ 2015, \aap, 584, A16 

\bibitem[Pinilla et al.(2016a)]{pinilla2016a} Pinilla, P., Klarmann, L., Birnstiel, T., et al.\ 2016a, \aap, 585, A35 

\bibitem[Pinilla et al.(2016b)]{pinilla2016b} Pinilla, P., Flock, M., Ovelar, M.~d.~J., \& Birnstiel, T.\ 2016b, \aap, 596, A81 

\bibitem[Pohl et al.(2016)]{pohl2016} Pohl, A., Kataoka, A., Pinilla, P., et al.\ 2016, \aap, 593, A12  

\bibitem[Reg{\'a}ly et al.(2012)]{regaly2012} Reg{\'a}ly, Z., Juh{\'a}sz, A., S{\'a}ndor, Z., \& Dullemond, C.~P.\ 2012, \mnras, 419, 1701 

\bibitem[Reipurth \& Zinnecker(1993)]{reipurth1993} Reipurth, B., \& Zinnecker, H.\ 1993, \aap, 278, 81 

\bibitem[Ricci et al.(2010)]{ricci2010} Ricci, L., Testi, L., Natta, A., \& Brooks, K.~J.\ 2010, \aap, 521, A66

\bibitem[Rosotti et al.(2013)]{rosotti2013} Rosotti, G.~P., Ercolano, B., Owen, J.~E., \& Armitage, P.~J.\ 2013, \mnras, 430, 1392 

\bibitem[Rosotti et al.(2016)]{rosotti2016} Rosotti, G.~P., Juhasz, A., Booth, R.~A., \& Clarke, C.~J.\ 2016, \mnras, 459, 2790 

\bibitem[Simon et al.(1995)]{simon1995} Simon, M., Ghez, A.~M., Leinert, C., et al.\ 1995, \apj, 443, 625 

\bibitem[Sokal(1994)]{sokal1994} Sokal, A.~D.\ 1994, arXiv:hep-lat/9405016 

\bibitem[Stanke \& Zinnecker(2000)]{stanke2000} Stanke, T., \& Zinnecker, H.\ 2000, IAU Symposium, 200, 38 

\bibitem[Stolker et al.(2016)]{stolker2016} Stolker, T., Dominik, C., Avenhaus, H., et al.\ 2016, \aap, 595, A113 

\bibitem[Strom et al.(1989)]{strom1989} Strom, K.~M., Strom, 
S.~E., Edwards, S., Cabrit, S., \& Skrutskie, M.~F.\ 1989, \aj, 97, 1451 

\bibitem[Tazzari et al.(2016)]{tazzari2016} Tazzari, M., Testi, L., Ercolano, B., et al.\ 2016, \aap, 588, A53 

\bibitem[Testi et al.(2014)]{testi2014} Testi, L., Birnstiel, T.,
  Ricci, L., et al.\ 2014, Protostars and Planets VI, 339,
  eds. H. Beuther, R. Klessen, C. Dullemond, \& Th. Henning (Univ. of Arizona Press, Tucson).

\bibitem[Trotta et al.(2013)]{trotta2013} Trotta, F., Testi, L., Natta, A., Isella, A., \& Ricci, L.\ 2013, \aap, 558, A64 

\bibitem[van der Marel et al.(2013)]{marel2013} van der Marel, N., van Dishoeck, E.~F., Bruderer, S., et al.\ 2013, Science, 340, 1199 

\bibitem[van der Marel et al.(2015)]{marel2015} van der Marel, N., van Dishoeck, E.~F., Bruderer, S., P{\'e}rez, L., \& Isella, A.\ 2015, \aap, 579, A106

\bibitem[van der Marel et al.(2015)]{marel2015_irs48} van der Marel, N., Pinilla, P., Tobin, J., et al.\ 2015, \apjl, 810, L7 

\bibitem[van der Marel et 
al.(2016)]{marel2016} van der Marel, N., van Dishoeck, E.~F., Bruderer, S., et al.\ 2016, \aap, 585, A58 

\bibitem[Walsh et al.(2016)]{walsh2016} Walsh, C., Juh{\'a}sz, A., Meeus, G., et al.\ 2016, \apj, 831, 200 

\bibitem[Whipple(1972)]{whipple1972} Whipple, F.~L. 1972, From Plasma to Planet, 211

\bibitem[White et al.(2016)]{white2016} White, J.~A., Boley, A.~C., Dent, W.~R.~F., Ford, E.~B., \& Corder, S.\ 2016, arXiv:1612.01648 

\bibitem[Wilking et al.(1989)]{wilking1989} Wilking, B.~A., Lada, C.~J., \& Young, E.~T.\ 1989, \apj, 340, 823 

\bibitem[Wilking et al.(2005)]{wilking2005} Wilking, B.~A., Meyer, M.~R., Robinson, J.~G., \& Greene, T.~P.\ 2005, \aj, 130, 1733 

\bibitem[Williams \& Best(2014)]{williams2014} Williams, J.~P., \&
  Best, W.~M.~J.\ 2014, \apj, 788, 59 

\bibitem[Zhang et al.(2016)]{zhang2016} Zhang, K., Bergin, E.~A., Blake, G.~A., et al.\ 2016, \apjl, 818, L16 

\bibitem[Zhu et al.(2012)]{zhu2012} Zhu, Z., Nelson, R.~P., Dong, R., Espaillat, C., \& Hartmann, L.\ 2012, \apj, 755, 6

\bibitem[Zsom et al.(2011)]{zsom2011} Zsom, A., S{\'a}ndor, Z., \& Dullemond, C.~P.\ 2011, \aap, 527, A10 

\end{thebibliography}
\end{document}